\documentclass[12pt]{article}
\pdfoutput=1

\usepackage[tracking=true,expansion=true,stretch=15,shrink=15]{microtype}
\DeclareMicrotypeSet*[tracking]{my}
{ font = */*/*/sc/* } 
\SetTracking{ encoding = *, shape = sc }{ 45 }
\SetProtrusion{encoding={*},family={bch},series={*},size={6,7}}
{1={ ,750},2={ ,500},3={ ,500},4={ ,500},5={ ,500},
	6={ ,500},7={ ,600},8={ ,500},9={ ,500},0={ ,500}}
\SetExtraKerning[unit=space]
{encoding={*}, family={qhv}, series={b}, size={large,Large}}
{1={-200,-200}, 
	\textendash={400,400}}

\usepackage[authoryear]{natbib}

\usepackage[pagewise]{lineno}
\usepackage{sectsty}
\usepackage{setspace}
\sectionfont{\centering}

\usepackage[latin1]{inputenc}
\usepackage{url}
\usepackage{amsmath,amsfonts,amssymb,amscd,amsthm,bm,bbm}
\usepackage{multirow}
\usepackage{dsfont}
\usepackage{mathrsfs}
\usepackage{empheq}
\usepackage{hyperref}
\usepackage{nameref}
\def\equationautorefname~#1\null{Equation~(#1)\null}
\usepackage{algorithm}
\usepackage{algpseudocode}

\hypersetup{
	colorlinks   = true, 
	urlcolor     = blue!50, 
	linkcolor    = blue!50, 
	citecolor   = blue!50 
}

\newcommand{\greekvec}[1]{\mbox{\boldmath$#1$}}

\newcommand{\vtheta}{\greekvec{\theta}}
\newcommand{\vgamma}{\greekvec{\gamma}}
\newcommand{\vbeta}{\greekvec{\beta}}

\newcommand{\bs}{\mathbf{s}}

\newcommand{\bd}{\mathbf{d}}

\newcommand{\bw}{\mathbf{w}}

\newcommand{\bx}{\mathbf{x}}

\newcommand{\bu}{\mathbf{u}}

\newcommand{\bp}{\mathbf{p}}
\newcommand{\ba}{\mathbf{a}}

\newcommand{\pmle}{\widehat{\vtheta}}

\newcommand{\ind}{\mathds{1}}
\newcommand{\R}{\mathds{R}}

\newtheoremstyle{new}{12pt}{12pt}{\itshape}{}{\bfseries}{.}{1em}{}
\theoremstyle{new}

\newtheorem{Example}{Example}

\theoremstyle{remark}

\usepackage{tikz}
\usepackage{subfig}
\usetikzlibrary{shapes,arrows,automata,positioning}

\tikzstyle{nn1}=[ellipse,font=\small,minimum width=1.2cm,draw=black, text=black,align=center]
\tikzstyle{ee1}=[-,font=\small,draw = black, text=black,sloped, above]

\algnewcommand{\Inputs}[1]{%
  \State \textbf{Inputs:}
  \Statex \hspace*{\algorithmicindent}\parbox[t]{.8\linewidth}{\raggedright #1}
}
\algnewcommand{\Initialize}[1]{%
  \State \textbf{Initialize:}
  \Statex \hspace*{\algorithmicindent}\parbox[t]{.8\linewidth}{\raggedright #1}
}

\makeatletter

\newcommand\Autoref[1]{\@first@ref#1,@}
\def\@throw@dot#1.#2@{#1}
\def\@set@refname#1{
    \edef\@tmp{\getrefbykeydefault{#1}{anchor}{}}%
    \def\@refname{\@nameuse{\expandafter\@throw@dot\@tmp.@autorefname}s}%
}
\def\@first@ref#1,#2{%
  \ifx#2@\autoref{#1}\let\@nextref\@gobble
  \else%
    \@set@refname{#1}
    \@refname~\ref{#1}
    \let\@nextref\@next@ref
  \fi%
  \@nextref#2%
}
\def\@next@ref#1,#2{%
   \ifx#2@ and~\ref{#1}\let\@nextref\@gobble
   \else, \ref{#1}
   \fi%
   \@nextref#2%
}

\makeatother

\newcommand{\blind}{0}

\begin{document}
\def\spacingset#1{\renewcommand{\baselinestretch}%
{#1}\small\normalsize} \spacingset{1}

\if0\blind
{
  \title{\bf Generalized Additive Models for Pair-Copula Constructions}
  \author{Thibault Vatter\thanks{Thibault Vatter is Postdoctoral researcher, Department of Statistics, 1255 Amsterdam Avenue, MC 4690, Columbia University, New York, NY 10027, USA (email: tv2233@columbia.edu)}\hspace{.2cm}\\
    Department of Statistics, \\
	Columbia University, \\
	New York, USA \\
    and \\
    Thomas Nagler\thanks{Thomas Nagler is PhD candidate, Lehrstuhl f\"ur Mathematische Statistik, Technische Universit\"at M\"unchen, Boltzmannstra\ss e 3,
85748 Garching b. M\"unchen, Germany (email: thomas.nagler@tum.de)}\hspace{.2cm} \\
    Lehrstuhl f\"ur Mathematische Statistik, \\
Technische Universit\"at M\"unchen, \\Munich, Germany}
  \maketitle
} \fi

\if1\blind
{
  \bigskip
  \bigskip
  \bigskip
  \begin{center}
    {\LARGE\bf Generalized Additive Models for Pair-Copula Constructions}
\end{center}
  \medskip
} \fi

\bigskip
\begin{abstract}
Pair-copula constructions are flexible dependence models that use bivariate copulas as building blocks. In this paper, we use generalized additive models to extend them by allowing covariates effects. Borrowing ideas from a traditionally univariate context, we let each pair-copula parameter depend directly on the covariates in a parametric, semiparametric or nonparametric way. We propose a sequential estimation method that we study by simulation, and apply it to investigate the time-varying dependence structure between the intraday returns on four major foreign exchange rates. An R package, a script reproducing the results in this article, and additional simulation results are provided as supplementary material.
\end{abstract}

\noindent%
\vfill

\newpage

\section{Introduction}
\label{sec:intro}

\textit{Pair-copula constructions} (PCCs) are flexible representations of the dependence underlying a multivariate distribution. Popularized in \cite{Bedford01, Bedford02, aasczadofrigessibakken2009}, they have become a hot topic of multivariate analysis over the last couple of years. The idea is to model the joint distribution of a $d$-dimensional random vector by considering pairs of conditional random variables. Let us consider a three dimensional example. The joint density $f_{1,2,3}(\bm x), \bm x \in \mathbb{R}^3,$ of a vector of continuous random variables $\bm X = (X_1, X_2, X_3)$ can be decomposed as

\begin{align*}
	f_{1,2,3}(\bm x)  =& f_1(x_1) \times f_2(x_2) \times f_3(x_3) \times c_{1,2}\left\{F_1(x_1), F_2(x_2)\right\} \times c_{2,3}\left\{F_2(x_2), F_3(x_3)\right\} \\
	&\times c_{1,3; 2}\left\{F_{1|2} (x_1 \mid x_2), F_{3|2} (x_3 \mid x_2); x_2\right\},
\end{align*}
where 
\begin{itemize}
\item $f_1, f_2, f_3$ (and $F_1, F_2, F_3$) are the marginal densities (and distributions), 
\item $c_{1, 2}$ is the joint density of $F_1(X_1)$ and $F_2(X_2)$, 
\item $c_{2, 3}$ is the joint density of $F_2(X_2)$ and $F_3(X_3)$, 
\item $c_{1, 3; 2}$ is the joint density of $F_{1|2} (X_1 \mid X_2)$ and $F_{3|2} (X_3 \mid X_2)$ conditional on $ X_2 = x_2$.
\end{itemize}
The above decomposition can be generalized to an arbitrary dimension $d$ and leads to tractable and very flexible models.

In general, the conditional density $c_{1, 3; 2}$ is also a function of $x_2$. However, this effect is often ignored for the sake of tractability, in which case we speak about a simplified PCC. When this so-called \emph{simplifying assumption} is made, the complete joint distribution can be built using unconditional bivariate copulas. Discussions on the simplifying assumption can be found in \citet{hobaekaasfrigessi2010}, \citet{Stoeber13}, and \citet{Spanhel15}.

An natural extension of PCCs includes the effect of covariates. This is particularly useful when one wants to investigate the influence of exogenous variables (such as space or time) on a complex dependence structure. For instance, the joint spatio-temporal modeling of several hydrograph flood variables, such as the flood peak, the hydrograph volume and hydrograph duration, is necessary to design and manage risks for hydraulic structures like dams \citep{requenamedierogarrote2013}.
Another example is the modeling of the joint distribution of intraday returns on exchange rates, whose the dependence structure changes over time due to the cyclical nature of market activity. Even when the covariate is random, we are often only willing to study its effect on the joint distribution of a response vector of interest. In this case, it is usually unnecessary or inconvenient to model its stochastic behavior explicitly, and a regression-like theory for PCCs is required. In the hydrological example above, when a region under study characterized by large hydro-climatic heterogeneities, the inclusion of additional (potentially random) descriptors in the model is important; especially as the ultimate goal is the extrapolation (prediction) at ungauged sites. Similarly, scheduled economic news, such as the monthly release of the US unemployment rate, or the Federal Open Market Committee (FOMC) press conference, impact in a crucial way the distribution of intraday returns \citep{andersen_bollerslev:1997,andersen_bollerslev:1998}. Previous work in this direction includes regime switching PCC \citep{stoeber-czado-ms}, and spatial PCC models \citep{Graeler14,Erhardt15bio, Erhardt15jmv}, where the individual parameters of the pair-copulas were modeled as linear functions of distances between different locations.

To relax the simplifying assumption and model the influence of covariates, the appropriate statistical tool is the \emph{conditional copula}. This made its first appearance in the seminal work of \citet{patton2002} in a time-series context. While Patton's approach is parametric, a fully nonparametric alternative was later proposed by \citet{gijbelsveraverbekeomelka2011} and \citet{Veraverbeke11}. \citet{acarcraiuyao2011} discuss a semiparametric model where the  dependence parameter is modeled as a smooth nonparametric function of a covariate, which is estimated by a kernel-based local likelihood approach. This methodology was used by  \citet{acargenestneslehova2012} for the estimation of the conditional dependence in a three-dimensional PCC.  A Bayesian method was proposed by \citet{craiusabeti2012} and extended by \citet{sabetiweicraiu2014} to allow for multiple covariates. 

Recently, \citet{vatterchavez2015} proposed an alternative approach based on generalized additive models (GAMs, see \citealt{hastietibshirani1990,greensilverman2000}) and spline smoothing. Building on the flexibility of GAMs, the copula parameter is modeled as a parametric, semiparametric or non-parametric function of the covariates. To maximize their (quadratically) penalized log-likelihood, \citet{vatterchavez2015} linearize a step of the Newton-Raphson algorithm, treat the approximation as Gaussian, find its solution, and iterate until convergence. As there exists mature software dealing with the Gaussian case  \citep{wood2004,wood2006,wood2011}, this method is stable and fast, even for large datasets \citep{Wood2015,Wood2016}.

All of those methods deal with inference for the conditional copula only. In other words, they use a two-step method called \emph{inference function for margins approach} \citep[IFM, see][]{joexu1996,joe2005} to estimate the margins first, and then the copula. Trading-off computational against statistical efficiency, another recent strand of research \citep{Klein2016,Radice2016,Marra2017} aims at estimating both in one-step, but for bivariate responses only. Because one-step estimation is hardly feasible when the dimension of the responses grows, we do not pursue this direction and use the IFM whenever marginal distributions are needed. 

In this paper, we use the method of \citet{vatterchavez2015} to model covariates effects for each pair-copula of a PCC in a parametric, semiparametric or nonparametric way. While the other methods mentioned above are restricted to bivariate responses, we are the first, to the best of our knowledge, to consider covariates effects on conditional copulas of larger dimension; the exception being \citet{acargenestneslehova2012}, who let a trivariate PCC be function of a single covariate.  

While \citet{vatterchavez2015} use the \texttt{gam} function from the R \citep{R} package \texttt{mgcv} \citep{wood2004,wood2006,wood2011}, their work is readily extensible: since most research on GAMs is developed and implemented for the Gaussian log-likelihood, the solution of the linearized Newton-Raphson step can be obtained using any suitable alternative instead of \texttt{mgcv}. For instance, sparsity-enforcing penalties \citep{Chouldechova2015,Lou2016,Petersen2016} or Boosting \citep{Buhlmann2003,Buhlmann2007,Tutz2007,Schmid2008} could be implemented to handle high-dimensional covariates.

Note that various tools to apply generalized additive models to bivariate copulas and PCCs are collected in an R package. Available on the Comprehensive R Archive Network at \url{https://cran.r-project.org/web/packages/gamCopula/}, \texttt{gamCopula} includes functions for parameter estimation, model selection, simulation, and visualization. 

The structure of the paper is as follows. In \autoref{sec:background}, we first introduce PCCs and the GAM framework of \citet{vatterchavez2015}. Then, we discuss inference issues related to PCCs with covariates. We study the estimator's behavior by simulation in \autoref{sec:simul}. In \autoref{sec:appli}, we model the time-varying dependence structure between the intraday returns on four major foreign exchange rates. We conclude with a discussion in \autoref{sec:discussion}.

\section{Methodology}
\label{sec:background}

\subsection{Pair-Copula Constructions}

\label{sec:pcc}
Let $\bm X = (X_1, \dots, X_d) \sim F$ be a  $d$-variate random vector. By the theorem of \citet{sklar1959}, any $F$ can be represented by its marginal distributions $F_1, \dots, F_d$ and a \emph{copula} $C$, which is is the joint distribution of $\bm U = (U_1, \dots, U_d) = \bigl(F_1(X_1), \dots, F_d(X_d)\bigr)$. If all distributions are differentiable, we can write

\begin{align*} 
    f(x_1, \dots, x_d) = c\bigl\{ F_1(x_1), \dots, F_d(x_d)\bigr\} \times \prod_{k=1}^d f_k(x_k),
\end{align*}
where $f,c, f_1, \dots, f_d$ are the densities corresponding to $F,C, F_1, \dots, F_d$ respectively. 

In this context, any $c$ can be decomposed into a product of $d(d-1)/2$ bivariate copula densities \citep{Joe97, Bedford01, Bedford02}. While a decomposition is not unique, it can be organized as a graphical model called \emph{regular vine (R-vine)}, namely a sequence of trees $T_m = (V_m, E_m)$ $(m = 1, \dots, d-1)$ also called the \emph{structure} of the PCC. Identifying each edge $e \in E_m$ with a bivariate copula $c_{j_e, k_e ; D_e}$, the joint density can then be written as the product of all pair-copula densities:

\begin{align}
c(\bm u) = \prod_{m=1}^{d-1} \prod_{e \in E_m} c_{j_e, k_e; D_e} \left\{ u_{j_e|D_e}, u_{k_e|D_e} ; \, \bm u_{D_e} \right\}, \label{sec:gampcc:density_nonsimplified_eq}
\end{align}
where $u_{j_e|D_e} := C_{j_e|D_e}(u_{j_e}\mid \bm u_{D_e})$, $\bm u_{D_e}:=(u_\ell)_{\ell \in D_e}$ is a subvector of $\bm u =(u_1, \dots, u_d) \in [0,1]^d$ and $C_{j_e|D_e}$ is the conditional distribution of $U_{j_e} \mid \bm U_{D_e}$. The set $D_e$ and the indices $j_e, k_e$ form respecively the \emph{conditioning set} and the \emph{conditioned set}. Put differently, $c_{j_e, k_e; D_e}$ describes the dependence between $U_{j_e}$ and $U_{k_e}$, conditional on $\bm U_{D_e}$.

\begin{Example}   \label{sec:gampcc:vine_ex}
	The density corresponding to the tree sequence in \autoref{sec:gampcc:RVine_fig} is
	
	\begin{align*}
	c(u_1, \dots, u_5) &= c_{1,2}(u_1, u_2) \times c_{1,3}(u_1,u_3) \times c_{3,4}(u_3,u_4) \times c_{3,5}(u_3,u_5)\\
	& \phantom{=} \times c_{2,3;1}(u_{2|1}, u_{3|1}; u_1) \times c_{1,4;3}(u_{1|3}, u_{4|3}; u_3) \times c_{1,5;3}(u_{1|3}, u_{5|3}; u_3) \\
	& \phantom{=} \times c_{2,4;1,3}(u_{2|1,3}, u_{4|1,3}; \bm u_{\{1,3\}}) \times c_{4,5;1,3}(u_{4|1,3}, u_{5|1,3}; \bm u_{\{1,3\}}) \\
	& \phantom{=} \times c_{2,5;1,3,4}(u_{2|1,3,4}, u_{5|1,3,4}; \bm u_{\{1,3,4\}}).
	\end{align*}
\end{Example}
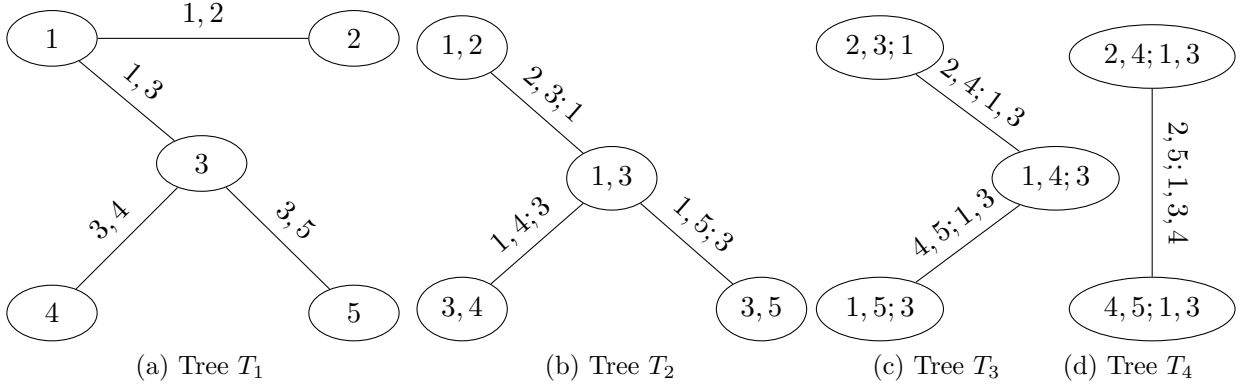
\begin{figure}
	\centering
\subfloat[Tree $T_1$]{
\begin{minipage}[t]{.32\linewidth}
     \centering
\begin{tikzpicture}[node distance=2cm,>=latex']
\node [nn1] (U1) {$1$};
\node [nn1, below right =1.6cm of U1] (U3) {$3$};
\draw [ee1] (U1) --  node {$1,3$} (U3) ;
\node [nn1, right = 2.8 cm of U1] (U2) {$2$};
\draw [ee1] (U1) --  node {$1,2$} (U2) ;
\node [nn1, below  = 2.9cm of U1] (U4) {$4$};
\draw [ee1] (U3) --  node {$3,4$} (U4) ;
\node [nn1, right = 2.8cm of U4] (U5) {$5$};
\draw [ee1] (U3) --  node {$3,5$} (U5) ;
\end{tikzpicture}
\end{minipage}
} \label{fig:rvineex1}
\hspace*{-0.3cm}
\subfloat[Tree $T_2$]{
\begin{minipage}[t]{.32\linewidth}
     \centering
\begin{tikzpicture}[node distance=2cm,>=latex']
\node [nn1] (U1) {$1,2$};
\node [nn1, below right =1.6cm of U1] (U3) {$1,3$};
\draw [ee1] (U1) --  node {$2,3;1$} (U3) ;
\node [nn1, below left =1.6cm of U3] (U4) {$3,4$};
\draw [ee1] (U3) --  node {$1,4;3$} (U4) ;
\node [nn1, below right =1.6cm of U3] (U5) {$3,5$};
\draw [ee1] (U3) --  node {$1,5;3$} (U5) ;
\end{tikzpicture}
\end{minipage} 
\label{fig:rvineex2}
} 
\hspace*{-0.3cm}
\subfloat[Tree $T_3$]{
\begin{minipage}[t]{.2\linewidth}
     \centering
\begin{tikzpicture}[node distance=2cm,>=latex']
\node [nn1] (U1) {$2,3;1$};
\node [nn1, below right =1.6cm of U1] (U3) {$1,4;3$};
\draw [ee1] (U1) --  node {$2,4;1,3$} (U3) ;
\node [nn1, below left =1.6cm of U3] (U4) {$1,5;3$};
\draw [ee1] (U3) --  node {$4,5;1,3$} (U4) ;
\end{tikzpicture}
\end{minipage}
\label{fig:rvineex3}
} 
\hspace*{-0.3cm}
\subfloat[Tree $T_4$]{
\begin{minipage}[t]{.1\linewidth}
     \centering
\begin{tikzpicture}[node distance=2cm,>=latex']
\node [nn1] (U1) {$2,4;1,3$};
\node [nn1, below =2.5cm of U1] (U3) {$4,5;1,3$};
\draw [ee1] (U1) --  node {$2,5;1,3,4$} (U3) ;
\end{tikzpicture}
\end{minipage}
\label{fig:rvineex4}
}
	\caption{Example of a Regular Vine Tree Sequence. The numbers represent the variables, $x,y$ denote the bivariate distribution of $x$ and $y$, and $x,y;z$ denote the bivariate distribution of $x$ and $y$ conditional on $z$. Each edge corresponds to a bivariate pair-copula in the PCC.}
	\label{sec:gampcc:RVine_fig}
\end{figure}
Because the conditioning set grows with the tree level, it is often convenient to ignore the influence of $\bm u_D{_e}$ on the pair-copula density $c_{j_e, k_e; D_e} $. Thus, the density of a \emph{simplified PCC} collapses to

\begin{align}
c(\bm u) = \prod_{m=1}^{d-1} \prod_{e \in E_m} c_{j_e, k_e; D_e} \left\{ u_{j_e|D_e}, u_{k_e|D_e} \right\}. \label{sec:gampcc:density_simplified_eq}
\end{align}
However, since each pair-copula can be modeled separately, simplified PCCs are highly flexible. Furthermore, they can easily be interpreted, as each pair-copula describes the dependence for a specific (conditional) bivariate random vector. For a more extensive treatment, we refer to \citet{aasczadofrigessibakken2009} and \cite{czado2010}.

\subsection{Generalized Additive Models for Pair-Copula Constructions}
\label{sec:gampcc}

To both relax the simplifying assumption and include a vector $\bm w$ of exogenous covariates, we need to model the conditional copula density for each edge of the PCC explicitly. The joint density can thus be rewritten

\begin{align}
c(\bm u; \bm w) = \prod_{m=1}^{d-1} \prod_{e \in E_m} c_{j_e, k_e; D_e} \bigl\{C_{j_e|D_e}(u_{j_e} \mid \bm u_{D_e}), C_{k_e|D_e}(u_{k_e} \mid \bm u_{D_e}); \, \bm w_e \bigr\}, \label{sec:gampcc:density_exo_eq}
\end{align}
where $\bm w_e  = (\bm w, \bm u_{D_e})$ contains both the exogenous covariates and the variables in the conditioning set. Note that $\bm w_e  = \bm w$ for $e \in E_1$.  Furthermore, if $\bm w_e = \bm u_{D_e}$ for each $e \in E_m$ ($m = 1, \dots, d-1$), we have the non-simplified PCC from \autoref{sec:gampcc:density_nonsimplified_eq} with no exogenous covariates. Because, to the best of our knowledge, the only dataset known to violate the simplifying assumption was studied by \citet{acargenestneslehova2012}, the usefulness of non-simplified PCCs has yet to be established. Thus, while emphasizing that they can be handled by our methodology, we will not pursue this path any further in this paper. 

An important subclass of \autoref{sec:gampcc:density_exo_eq}, and the focus of the remainder of this paper, is obtained by writing $\bm w_e  = \bm w$ for each $e \in E_m$, $m = 1, \dots, d-1$. In this case, the simplifying assumption is satisfied, but the effects of $\bm w$ are still included, and we call this model a \emph{simplified PCC with exogenous covariates}. 

To make this construction practically useful, one assumes a parametric form for the (conditional) copula density $c_e\left\{\cdot , \cdot\, ; \eta_e(\bm w) \right\}$. For common copula families, there is a one-to-one mapping between the copula parameter $\eta_e$ and Kendall's $\tau_e$, a copula-based measure of concordance \citep[see, e.g.,][Section 5.1.1]{Nelsen06}. For simplicity, we assume throughout that such a relationship exists. When this is not the case (or for multi-parameter families), the methodology can still be applied to the copula parameter (or to one of them). See \citet{vatterchavez2015} for an additional discussion.

We can then reparameterize the conditional copula as a function of its corresponding Kendall's $\tau$, and we write $c_e(\cdot , \cdot\, ; \tau_e(\bm w))$.  Dropping the subscript $e$ for clarity, \citet{vatterchavez2015} proposed to model the variation of the dependence parameter with respect to the covariates as 

\begin{align}
\tau(\bm w; \vtheta) =  g\left\{\mathbf{z}^{\top}\greekvec{\beta} + \sum_{k=1}^{K} s_{k}(\mathbf{t}_{k})\right\},
\label{eq:condsemipara}
\end{align}
where:
\begin{itemize}
\item $g(x) = (e^{x}-1)/(e^x+1)$, namely the inverse Fisher z-transform, is the link between the GAM and Kendall's $\tau$, 
\item $\mathbf{z}$ and $\mathbf{t}_k$ are subsets of $\bm w$ or products thereof (to allow interactions),
\item $\greekvec{\beta} \in \R^{P}$ is a vector of parameters,
\item $s_{k}: \mathbb{S}_{k} \rightarrow \R $ are smooth functions supported on closed interval $\mathbb{S}_{k} \subset \R$ for all $k$, 
\item $\vtheta$ is the vector of stacked parameters, containing both $\greekvec{\beta}$ and $s_{k}$ for all $k$.
\end{itemize}
Models of this form are also called \emph{partially linear models} \citep{Haerdle07}: they consist of a linear component, $\mathbf{z}^{\top}\greekvec{\beta}$, and a nonlinear component, $\sum_{k=1}^{K} s_{k}(\mathbf{t}_{k})$.  Note that, as in \citet{vatterchavez2015}, any strictly increasing and infinitely continuously differentiable $g: \mathbb{R} \longmapsto [-1,1]$ could be used as a link instead. However, the Fisher z-transform is the closest there is to a \emph{canonical link} in this context, and we assume the link to be correctly specified. While out of the scope of this paper, we refer to the vast literature on link misspecification for more details on this. For instance, \citet{Li1989} provide an early analysis of link misspecification, \citet{Czado2000} study thoroughly via simulations the effects of switching to noncanonical links, and \citet{Horowitz2001} suggest a nonparametric estimation method when the link is unknown.

\begin{figure}
    \centering
	\centering
\subfloat[Tree $T_1$]{
\begin{minipage}[t]{.45\linewidth}
     \centering
\begin{tikzpicture}[node distance=2cm,>=latex']
\node [nn1] (U1) {$1$};
\node [nn1, left =1cm of U1] (U2) {$2$};
\draw [ee1] (U1) --  node {$1,2$} (U2) ;
\node [nn1, right = 1cm of U1] (U3) {$3$};
\draw [ee1] (U1) --  node {$1,3$} (U3) ;
\end{tikzpicture}
\end{minipage} 
\label{fig:exogenex1}
} 
\hspace{0.5cm}
\subfloat[Tree $T_2$]{
\begin{minipage}[t]{.45\linewidth}
     \centering
\begin{tikzpicture}[node distance=2cm,>=latex']
\node [nn1] (U1) {$1,2$};
\node [nn1, right =1.5cm of U1] (U2) {$1,3$};
\draw [ee1] (U1) --  node {$2,3;1$} (U2) ;
\end{tikzpicture}
\end{minipage}
\label{fig:exogenex2}
} 
	\caption{A Three-Dimensional Regular Vine Tree Sequence. The numbers represent the variables, $x,y$ denote the bivariate distribution of $x$ and $y$, and $x,y;z$ denote the bivariate distribution of $x$ and $y$ conditional on $z$. Each edge corresponds to a bivariate pair-copula in the PCC.}
	\label{sec:gampcc:exogenex-fig}
\end{figure}
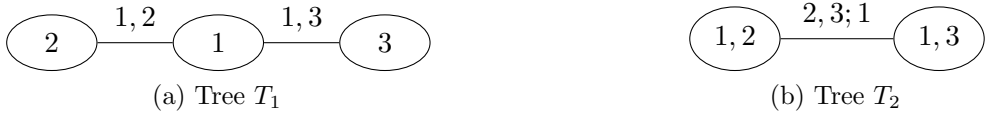

As an example, consider the three dimensional PCC from \autoref{sec:gampcc:exogenex-fig}. Furthermore, assume that it is simplified and that each pair-copula depends on a covariate $x$ and on time $t$. As such, the vector of covariates $\bm w = (x,t)$ is the same in tree $T_1$ and tree $T_2$.  Supposing that we want to allow for a nonlinear effect of time-variation and treat the effect of the other covariate as linear, a model for the three conditional pair-copulas can be written as

\begin{align*}
\tau_{1,2}(\bm w) &=  g\left\{ x\beta_{1,2}+s_{_{1,2}}(t)\right\}, \\
\tau_{1,3}(\bm w) &=  g\left\{ x\beta_{1,3}+s_{_{1,3}}(t)\right\}, \\
\tau_{2,3;1}(\bm w) &=  g\left\{ x\beta_{2,3;1}+s_{_{2,3;1}}(t)\right\}.
\end{align*}
In the non-simplified case, the covariate vector for the third pair-copula would be augmented such that $\bm w_{2,3;1} = (\bm w, u_1)$.

In this paper, we assume that all smooth functions $s_{k}$ are twice continuously differentiable on their support $\mathbb{S}_{k}$, and admit a finite-dimensional basis-quadratic penalty representation \citep[cf.][]{wood2006}. In particular, we focus on \emph{natural cubic splines} (NCSs). A NCS $s: \mathbb{S} \rightarrow \mathbb{R}$ is defined with a fixed sequence of $m$ knots, $\mbox{inf }  \mathbb{S} = y_{0}<y_{1}<\dots<y_{m}<y_{m+1}=\mbox{sup }   \mathbb{S}$. It is linear on the two extreme intervals $[y_{0},y_{1}]$ and $[y_{m},y_{m+1}]$ and twice continuously differentiable on $\mathbb{S}$. As such, it can be parametrized using $\bm s \in \mathbb{R}^m$, and there exists a unique $m\times m$ symmetric matrix $\mathbf{S}$ of rank $m-2$ such that $\int_{\mathbb{S}}s''(t)^2 \,dt=\mathbf{s}^{\top}\mathbf{S}\mathbf{s}$. This matrix is fixed in the sense that it depends on the knots but not on $\mathbf{s}$.

Let $m_k$ denote the basis size of $s_k$; then $\vtheta \in \Theta \subseteq \mathbb{R}^l$, where $p=P+\sum^K_{k=1} m_k$. For $\bu \in [0,1]^2$ and $\vtheta \in \Theta$, we denote the log-likelihood function conditional on $\bw$ by
$ \ell(\bu, \bw;\vtheta) =  \log c \left\lbrace \bu;\tau(\bw; \vtheta) \right\rbrace $, assuming that the dependence parameter is sufficient to identify the copula. Then, considering a random sample of $n$ observations $\left\lbrace \bu^j, \bw^j \right\rbrace^n_{j=1}$, the log-likelihood is $\ell(\vtheta) = n^{-1}\sum^n_{j=1}\ell(\bu^j, \bw^{j};\vtheta)$. Because,  for arbitrarily chosen basis sizes, its maximizer is unlikely to yield smooth estimates of $s_1,\, \dots,\, s_K$, we add roughness penalties to each nonparametric component, and define the penalized log-likelihood

 \begin{align}
 \ell(\vtheta, \vgamma) &= \ell(\vtheta) - \frac{1}{2} \sum_{k=1}^K \vgamma_{k}\int_{\mathbb{S}_{k}} s_{k}^{''}(\mathbf{t}_k)^2d\mathbf{t}_k  
 = \ell(\vtheta) - \frac{1}{2}\vtheta^{\top} \bp(\vgamma) \vtheta,
 \label{eq:lp}
 \end{align}
 with $\vgamma \in (\R_{+} \cup \left\lbrace 0 \right\rbrace)^K$, a vector of smoothing parameters, and $\bp(\vgamma)$ is a $p\times p$ block diagonal matrix with $K+1$ blocks; the first $P \times P$ block is filled with zeros and the remaining $K$ are equal to $\vgamma_k \bm S_k$, where $\bm S_k$ is the quadratic penalty representation of $s_k$. We define the penalized maximum log-likelihood estimator as
 
 \begin{align} \pmle = \underset{\displaystyle \vtheta \in \Theta}{\mbox{argmax }} \ell(\vtheta, \vgamma).
 \label{eq:pmle}
 \end{align}
In \citet{vatterchavez2015}, it is shown that one step of Fisher's scoring algorithm can be approximated by a generalized ridge regression. In other words, $\pmle$ is found iteratively by solving

 \begin{align}
 \vtheta^{\left[l+1\right]}(\vgamma) = \underset{\displaystyle \vtheta \in \Theta}{\mbox{argmin}} \left\lbrace \|\mathbf{y}^{\left[l\right]} -  \bd^{\left[l\right]} \vtheta \|_{\displaystyle \ba^{\left[l\right]}}^2 + \|\vtheta \|_{\displaystyle \bp(\vgamma)}^2 \right\rbrace,
 \label{eq:weighted-ridge}
 \end{align}
 where $\| \mathbf{x} \|_{\displaystyle \mathbf{w}}^2 = \mathbf{x}^{\top} \mathbf{w} \mathbf{x}$ and $\mathbf{y}^{\left[l\right]}$, $\bd^{\left[l\right]}$ and $\ba^{\left[l\right]}$ depend only on the data, $\vtheta^{\left[l\right]}$ and the copula family. With $\bs^{\left[l\right]}(\vgamma)$ the so-called influence or hat matrix  at the $l$th iteration, defined such that $\bs^{\left[l\right]}(\vgamma)\mathbf{y}^{\left[l\right]}=\bd^{\left[l\right]} \vtheta^{\left[l+1\right]}(\vgamma)$, we can now define the \emph{effective or equivalent degrees of freedom} (EDF) at the $l$th iteration as
 
 \begin{align*}
 \mbox{EDF}^{\left[l\right]}(\vgamma) &= \mbox{tr}\left\lbrace \bs^{\left[l\right]}(\vgamma)\right\rbrace.
 \end{align*}
 For additional discussions and alternative definitions of the EDF in the context of exponential families, see \citet{hastietibshirani1990} or \citet{greensilverman2000}. To balance goodness-of-fit and dimensionality, \citet{vatterchavez2015} minimize the \emph{generalized cross-validation} sum of squares \citep{craven_wahba:1978}
 
 \begin{align*}
 \mbox{GCV}^{\left[l\right]}(\vgamma) = \frac{n^{-1}\|\mathbf{y}^{\left[l\right]} -  \bs^{\left[l\right]}(\vgamma) \mathbf{y}^{\left[l\right]} \|_{\displaystyle \ba^{\left[l\right]}}^2}{\left\{1-n^{-1}\mbox{EDF}^{\left[l\right]}(\vgamma) \right\}^2}
 \end{align*}
 at each generalized ridge iteration, and a model's EDF is the final $\mbox{EDF}^{\left[l\right]}(\vgamma)$, obtained at convergence.

\subsection{Generalized Additive Model Selection for a Single Family}
\label{subsec:ems_pc}
At this step, the goal is to select a GAM assuming a known pair-copula family. Generally speaking, there can be at most one unique smooth function $s_{e,k}$ for each covariate. But usually there is no prior knowledge of its shape. This raises two questions:
\begin{enumerate}
\item Which of the covariates should be deemed unimportant, treated parametrically or nonparametrically?
\item What is the appropriate basis size and corresponding smoothing parameter for each of the smooth functions?
\end{enumerate}
Below, we give heuristically motivated answers to these questions, and we summarize this method in \autoref{algo1} (in the appendix). However, it should be noted that the problems of feature selection, determining which selected features should be treated as linear and nonparametric, and finding a suitable basis size while estimating the corresponding smoothing parameter for each nonparametric feature, represent active research areas for GAMs. For instance, recent work \citep{Chouldechova2015,Lou2016} suggests methods to solve the first two problems using overlapping grouped lasso penalties. But it is concerned with Gaussian and Binomial regression only, and uses a fixed basis size and/or smoothing parameter. Another strand of research \citep{wood2011,wood:pya:safk:2016,Wood2017} suggests widely applicable methods to select the smoothing parameters. But it forgoes the problems of features and basis sizes selection. 

To answer the {\bfseries first question}, we first set the basis size for each component to ten (i.e. $m_k = 10$ for each $k$), which is at the same time big enough to detect obvious non-linear relations and small enough to be quickly estimated. Second, we use a variant of backward elimination, where we start with all the covariates, remove at each step the ones whose individual $p$-values are above a pre-specified level $\alpha$, re-estimate the model and iterate until all remaining covariates are significantly non-zero. Third, terms whose EDFs are ``close'' to one are treated as linear components in the next step of the backward elimination.

In addition to the usual issues related to step-wise selection methods, $p$-values for the smooth terms are necessarily approximate. As suggested in \citet{marrawood2012} and \citet{wood2013a,wood2013b} in the exponential family context, we compute individual $p$-values using a Wald test. To compute the test statistics, we use a covariance matrix that results from the Bayesian interpretation of GAMs. While there is no optimality result for the power, \citet{marrawood2012}, extending the analysis of \citet{nychka1988}, motivated the use of this covariance matrix by showing that the resulting intervals have better frequentist performance (power and size under the null) than those computed using a strictly frequentist approximation.

As for the {\bfseries second question}, once the set of covariates is selected, the basis size of the smooth components is usually not critical. The reason is similar to univariate GAMs. The upper limit on the degrees of freedom associated with a smoother is given by its basis size. But the actual degrees of freedom is controlled by the penalization, as the corresponding smoothing parameter is selected during the fitting. Hence, while the exact choice of the basis size is not critical, it should be large enough to approximate well the data's underlying features and small enough to maintain good computational efficiency. To achieve this trade-off, we do the following:
\begin{enumerate}
\item Start with a small basis size for each smooth component.
\item Fit the GAM.
\item Check which of the estimated EDFs are ``close'' to the upper limit. If this is the case, increase the corresponding basis sizes.
\item Iterate 2. and 3. until no further increase is required for any of the individual basis sizes.
\end{enumerate}
In other words, for each smoother, we start with a small basis size and increase the ``model capacity'' until there is ``enough''. Additionally, at each step, we enforce a ``maximal model capacity'' in two ways. First, we make sure that no individual basis size is greater than the sample size divided by thirty. Second, in case there are ties in the covariate corresponding to a given smooth, we ensure that its basis size is smaller than one half of the number of unique values. While not theoretically justified, keeping reasonably large ratios of number of observations and unique covariates to number of parameters are rules of thumb that we found useful in this context. Apart from model capacity considerations, it should be noted that we also observed numerous numerical instabilities when rules were not enforced. 

\subsection{Sequential Estimation of a Pair-Copula Construction}
\label{subsec:ems_pcc}

To estimate PCCs, it is common to follow a sequential approach \citep[see e.g.][]{aasczadofrigessibakken2009,Haff13,Nagler16}, which we outline below. Assume that $\bm u^i = (u_1^i, \dots, u_d^i)$ ($i = 1, \dots, n$) are observations from a pair-copula construction and the vine structure is known. Then, the pair-copulas of the first tree, $T_1$, can be easily estimated using the method described in the previous subsection. This is not as straightforward for trees $T_m$ with $m \ge 2$ since data from the densities $c_{j_e, d_e;D_e}$ are unobserved. However, we can sequentially construct pseudo-observations by an appropriate transformation of the data.

Define the $h$-functions \citep[cf.][]{aasczadofrigessibakken2009} corresponding to a pair-copula density $c_{j_e, k_e;D_e}(u, v; \cdot)$ as

\begin{align*}
	h_{j_e|k_e;D_e }(u \mid v; \, \cdot) = \int_0^u c_{j_e, k_e;D_e}(s, v; \,\cdot) ds, \quad 
    h_{k_e|j_e;D_E}(u \mid v;\,  \cdot) = \int_0^v c_{j_e, k_e;D_e}(u, s;\, \cdot) ds,
\end{align*}
for all  $(u,v) \in [0,1]^2$. The dot in the third argument represents one of the GAM-formulations in Section \ref{sec:gampcc}. A crucial insight is the following: Assume we have (pseudo-)observations from the pair-copula density $c_{j_e,k_e;D_e}$, denoted as $(u_{j_e|D_e}^i, u_{k_e|D_e}^i)$ ($i = 1, \dots, n$). Then we can construct pseudo-observations for the next tree by setting

\begin{align}
	u_{j_e|D_e \cup k_e}^i = h_{j_e|k_e;D_e }\left(u_{j_e|D_e}^i \mid u_{k_e|D_e}^i;\, \cdot\right), \quad u_{k_e|D_e \cup j_e}^i = h_{k_e|j_e;D_e }\left(u_{k_e|D_e}^i \mid u_{j_e|D_e}^i;\, \cdot\right),\, i = 1, \dots, n. \label{sec:sequ:pobs_eq}
\end{align}

As only the estimates of each pair-copula in $T_l$ are  required to compute pseudo-observations for  tree $T_{l+1}$, we make use of the following sequential estimation and model selection procedure, starting with tree $T_1$: 
\begin{enumerate}
\item For each edge in the tree:
	\begin{enumerate}
		\item Select the covariates and estimate a GAM for each copula family via \autoref{subsec:ems_pc}. 
		\item Use the AIC to choose a copula family. 
		\item Use the estimates to construct pseudo-observations for the next tree via \eqref{sec:sequ:pobs_eq}.
	\end{enumerate}
\item Go to the next tree.
\end{enumerate}
The fact that the tree sequence $T_1, T_2, \dots,$ is a regular vine guarantees that at any step in this procedure, all required pseudo-observations are available. Note that the choice of the AIC as a trade-off between goodness of fit and model complexity is arbitrary. As such, any information criterion could be used instead and the BIC is also implemented the \texttt{gamCopula} package.
\section{Simulations}
\label{sec:simul}


\subsection{Setup} \label{sec:simul:setup}

We consider a simplified GAM-PCC in five dimensions using the vine structure depicted in \autoref{sec:gampcc:RVine_fig}. The Kendall's $\tau$ of each pair-copula is set as the partially linear model

\begin{align}
\tau_e(\bm z, t) = g\left\{\bm{z}^{\top}\greekvec{\beta} + \sum_{k = 1}^5 s_{e,k}(t_k)\right\}. \label{sim_gam}
\end{align}
The covariate vector $\bm z \in \mathbb{R}^{10}$ consists of five $\mathrm{Bernoulli}(0.5)$ variables and five standard normal variables (in that order). The covariates $t_k$ are independent standard uniform variables. In total there are 15 covariates, 10 for the linear component and 5 for the smooth component. We set 

\begin{align*}
\greekvec{\beta} = 1/4 \times (1, 1, -1, 0, 0, 1, -1, -1, 0, 0)^{\top},
\end{align*}
so that only six components of $\bm z$ actually have an influence on $\tau_e$. To encompass different cases of practical interest, we use the following deterministic functions:

\begin{align*}
s_{1}(t_1) =  -1/4 + t_1/2, \quad s_{2}(t_2) = \sin (2 \pi  t_2 )/4, \quad
s_{3}(t_3) = \sin (6 \pi  t_3 )/4, \quad s_4(t_4) = s_5(t_5) = 0,
\end{align*}
so that only the three covariates $t_1, t_2, t_3$ have an influence on $\tau_e$.
In order to make consistent estimation feasible, we represent each of the smooth functions in a cubic spline basis on 10 knots (equidistant on the unit interval).  

Additionally, we draw the copula family for each pair-copula with equal probability from the Gaussian, Student $t$ (with four degrees of freedom), Clayton and Gumbel families. The Clayton and Gumbel families are extended to allow for $\tau < 0$ by using $90^{\circ}$ and $270^{\circ}$ rotations. For example, if $c^{\mathrm{Clay}}(u_1, u_2; \tau)$ denotes the Clayton copula density with $\tau >0$, then $c^{\mathrm{Clay}90}(u_1, u_2; -\tau) = c(u_2,1 - u_1; \tau)$ is its 90 degree (counter-clockwise) rotation and allows for negative dependence. 

Finally, we repeat the experiment $500$ times for each of the two sample sizes $n=500, 5\,000$. Such sample sizes are common when modeling dependence in financial data. For instance, they represent between two and twenty years of daily financial data. While 500 observations may seem large for a ``small sample'' setup, it should be noted that the models under considerations contain $(4 + 3 + 2 + 1) \mbox{ (pair-copulas)} \times (1 + 6 +10\times 3) \mbox{ (linear and spline coefficients)} = 370$ parameters. Hence, with $n=500$ there are only slightly more observations than parameters to estimate.

 R code to reproduce all results is provided in the supplementary material.

\subsection{Results}

In what follows, we discuss the results of the simulation study outlined above. In the two-dimensional case, the penalized likelihood estimator was found to perform well by \citet{vatterchavez2015}. The pair-copulas from the first tree correspond to such a situation. In other trees, estimates are based on pseudo-observations, so estimation errors from the first tree (and subsequent ones) are expected to propagate and damage the performance. However, similarly as the first tree corresponding to the bivariate case, the dimension of the PCC is irrelevant to the performance for a given tree level. It should also be noted that high-dimensional PCCs are often truncated after the first few trees. As such, studying the effects of an increasing dimensionality is less relevant than the effects of the tree level. Hence, emphasis is put on how this performance changes with the tree level. 

We split the analysis in four parts. The first two parts discuss respectively the accuracy of estimates for the linear coefficient and the smooth functions. The third part concerns the selection of copula families and covariates. The fourth discusses computation times. We also investigated whether the copula family influences the estimation accuracy. Because the differences are tiny and do not lead to interesting insights, the results are not shown here.

\underline{Estimation of Linear Coefficients.}

\begin{figure}[!h]
\centering
\includegraphics[scale=0.8]{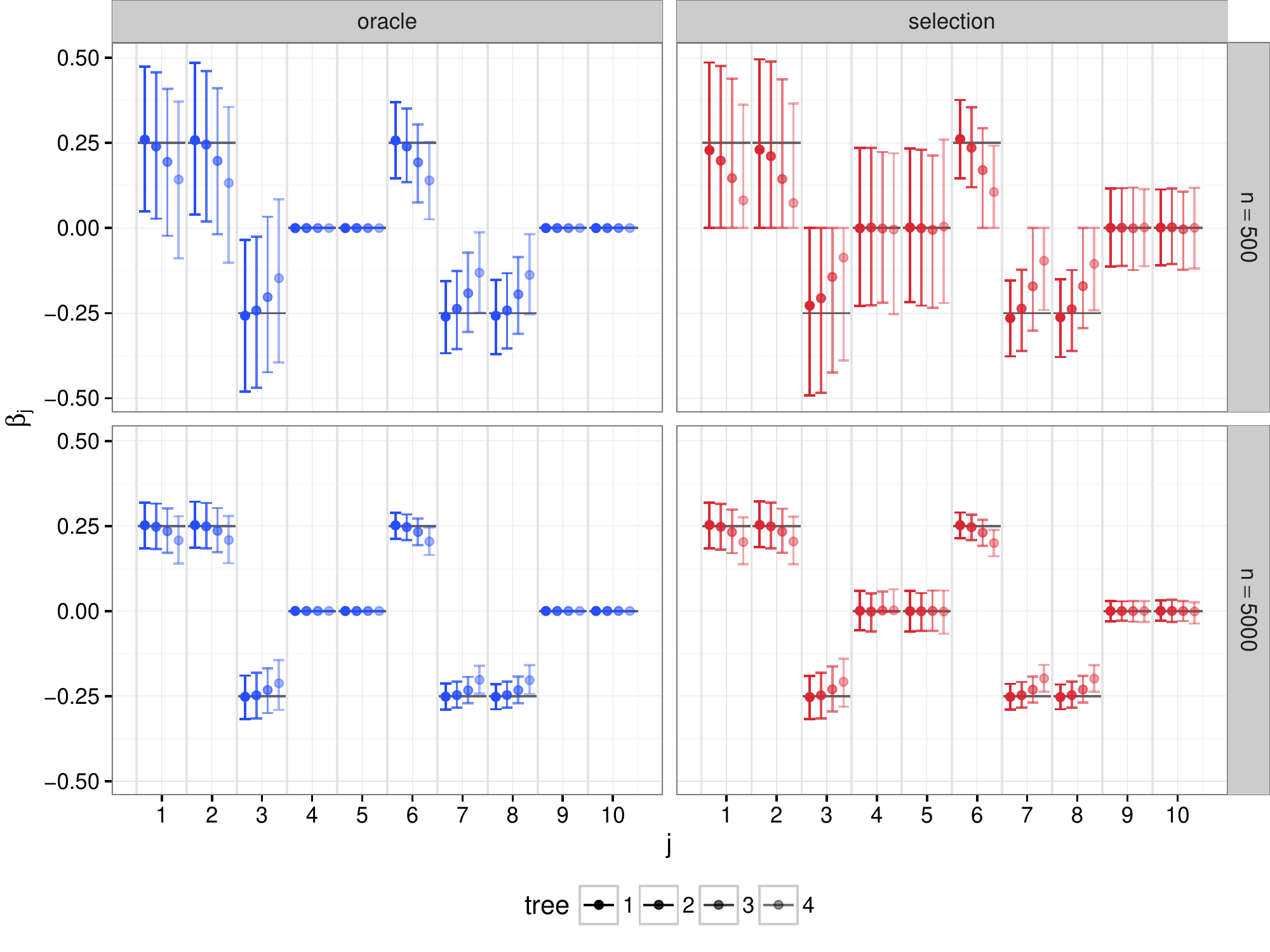}
\label{sec:simul:coefs_fig_5000}
\caption{Estimation of the Linear Coefficients. The estimates using the correctly specified estimator (oracle) and the model selection algorithm (selection) are compared. For each coefficient $\beta_j$ ($j = 1, \dots, 10$), the results are split by tree level (i.e., 1, 2, 3 or 4). Mean estimates are indicated by circles. Bars represent the range from the $5\%$ to the $95\%$ quantiles.} \label{sec:simul:coefs_fig}
\end{figure}

\noindent The accuracy of estimates of the linear coefficients $\greekvec \beta$ is illustrated in \autoref{sec:simul:coefs_fig}. The $x$-axis contains the ten entries of the vector $\greekvec \beta$. For each $\beta_j$ ($j = 1, \dots, 10$), \autoref{sec:simul:coefs_fig} shows four bars. Each bar represents the range from the $5\%$ to the $95\%$ quantiles of the 500 estimated coefficients, and the mean is shown as a circle. The four bars correspond to the first to fourth tree level of the PCC (from left to right). Horizontal bars indicate the true value of the coefficient. The left column corresponds to the oracle estimator where copula family and covariates included in the model are correctly specified. The right column corresponds to estimates resulting from the automatic model selection procedure described in \autoref{subsec:ems_pcc}.

For the coefficients $\beta_4$, $\beta_5$, $\beta_9$, $\beta_{10}$, the estimation error of the oracle estimator is zero, because the correct model specification does not include $z_4$, $z_5$, $z_9$, $z_{10}$. In contrast, the estimates from the model selection procedure fluctuate around zero, with a variance decreasing with the sample size. Interestingly, the tree level does not appear to affect the variance. For the other coefficients, namely $\beta_1$, $\beta_2$, $\beta_3$, $\beta_6$, $\beta_7$, $\beta_8$, the variance is similarly decreasing with the sample size, but unaffected by the tree level. 

The estimators are unbiased in the first tree, but in subsequent trees we observe a bias towards zero at all coefficients. The magnitude of this effect increases with the tree level and decreases with the sample size. This shrinkage towards zero is a consequence of the sequential procedure: estimation errors in a given tree propagate and disguise the effects of covariates in subsequent trees. In other words, the bias increases because pseudo-observations in subsequent trees are obtained from estimates of the previous tree(s). In parametric models with a large number of parameters, such shrinkage is often intentional to avoid overfitting (e.g., in the context of high-dimensional linear regression). In our case, the true model contains $370$ parameters. Even for a five-dimensional PCC model with only a single linear covariate, there are $20$ parameters to estimate. So although the shrinkage that we observe is not intentional, it may be opportune. 

The results also shed some light on the performance for PCCs with more than five variables. Since the pair-copulas are estimated tree after tree, the accuracy for the first four trees will be exactly the same as in \autoref{sec:simul:coefs_fig}, no matter how many variables are included in the PCC. Furthermore, the results show a clear trend going from the first tree to the fourth. We can expect this trend to continue when going to even higher tree levels.


\underline{Estimation of the Smooth Functions.}

\begin{figure}[t]
\centering
\includegraphics[width = \textwidth]{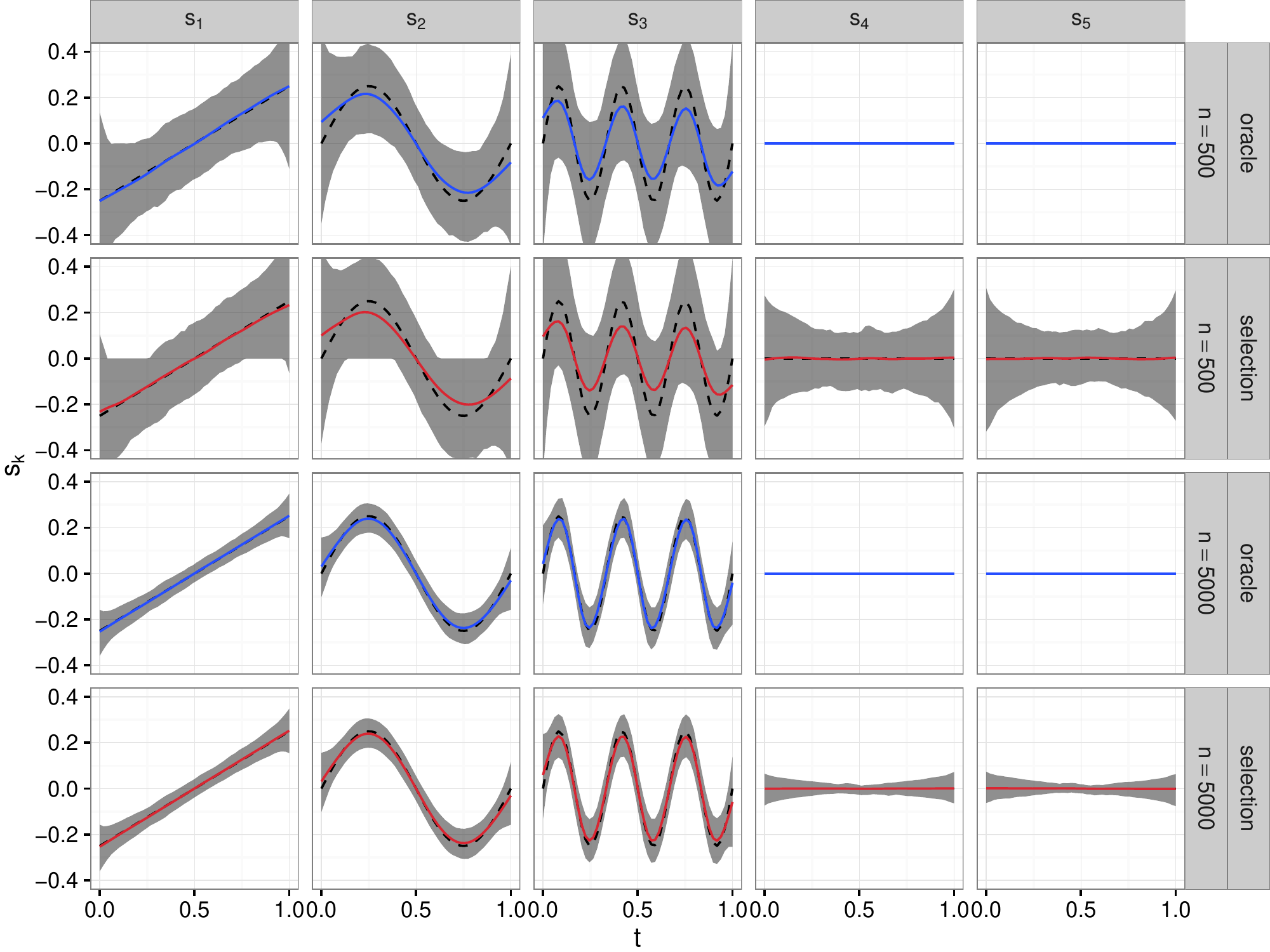}

\caption{Estimation of the smooth functions, tree $T_1$. The estimates using the correctly specified estimator (oracle) and the model selection algorithm (selection) are compared. The true calibration function (dashed line), the mean estimates (solid line) and pointwise range from the $5\%$ to the $95\%$ quantiles (shaded area).} \label{sec:simul:fundet_1}
\end{figure}

\noindent We now turn to the estimation of the smooth components $s_k$ ($k = 1, \dots, 5$), which are drawn as dashed lines in each of the columns of \autoref{sec:simul:fundet_1}. The first, $s_1$, is a simple linear function; $s_2$  and $s_3$ are sines with one and three wave periods.  They represent increasing complexity, and are expected to be increasingly difficult to estimate. The functions $s_4$ and $s_5$ are zero everywhere and therefore not included in the specification of the oracle estimator. Mean estimates are shown by solid lines with pointwise ranges from the $5\%$ to the $95\%$ quantiles as shaded areas. \autoref{sec:simul:fundet_1} shows the results for the first tree. Generally, the oracle and selection estimators show similar performance. An exception is the estimation of $s_4$ and $s_5$, where the selection estimator fluctuates around zero. For the functions $s_2$ and $s_3$ both estimators have a bias in regions where the functions display a high curvature. This is not surprising because a penalty is imposed on the second derivative of the fitted curves. This bias as decreases with the sample size and for $n = 5\,000$, both estimators are able to recover the functions accurately with only little variability.

\begin{figure}[t]
\includegraphics[width = \textwidth]{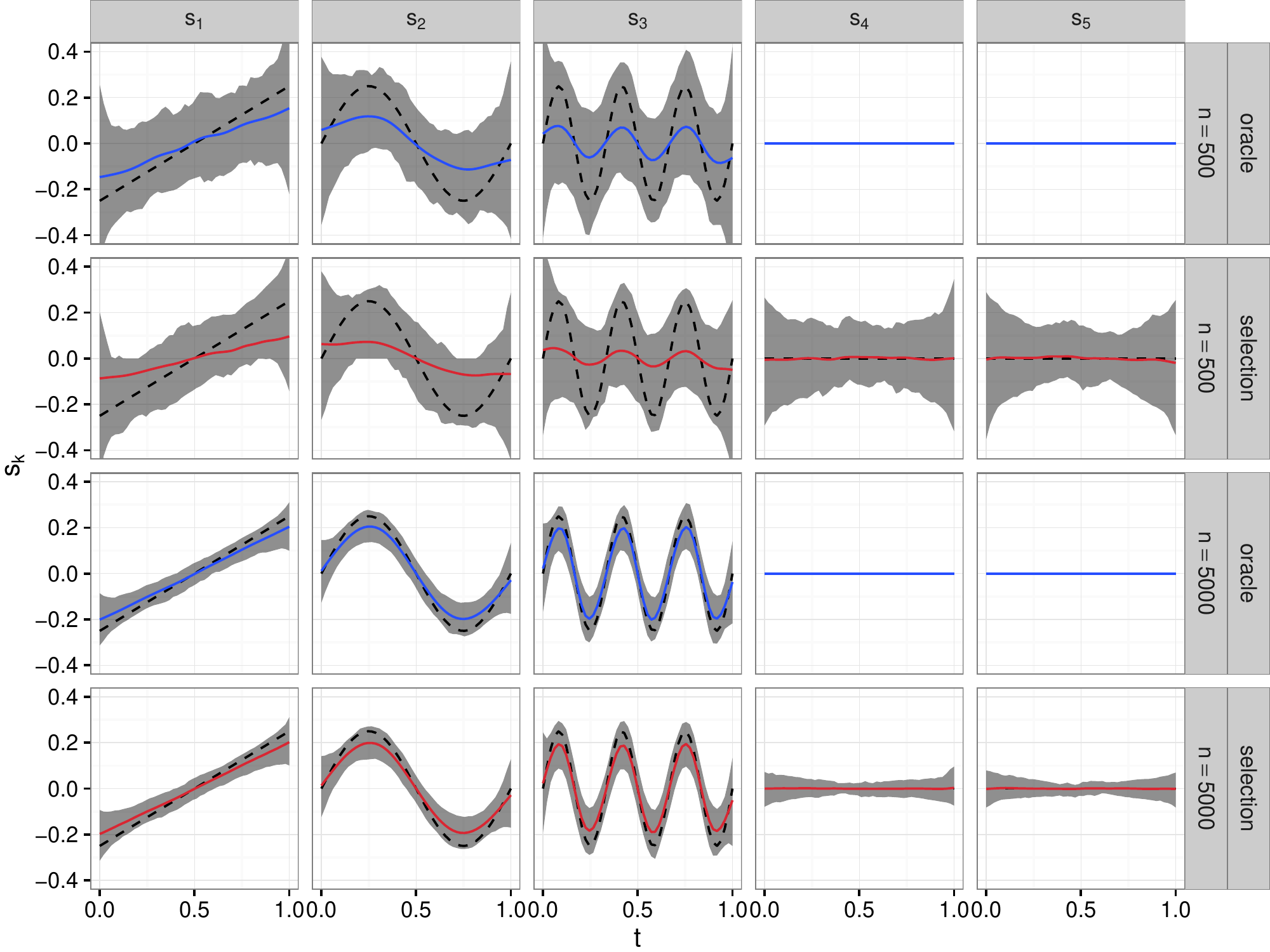}
\caption{Estimation of the smooth functions, tree $T_1$. The estimates using the correctly specified estimator (oracle) and the model selection algorithm (selection) are compared. The true calibration function (dashed line), the mean estimates (solid line) and pointwise range from the $5\%$ to the $95\%$ quantiles (shaded area).} \label{sec:simul:fundet_4}
\end{figure}

\autoref{sec:simul:fundet_4} shows the same story for the fourth tree. Similarly as for the linear coefficients, an additional bias towards zero appears. The bias is larger for the selection estimator, because the number of knots is also selected automatically. Often, fewer than ten knots are selected, and since the true function cannot be represented by such a basis, this causes additional bias. However, the bias is reduced substantially for $n =5\, 000$ for both the oracle and selection estimators.

Regarding the performance in PCC models with more than five variables, the statements made for estimating linear coefficients apply for nonlinear terms as well. 

\underline{Family Selection.}

\noindent \autoref{sec:simul:model_selection_1} shows the frequency that the true family is selected, split by sample size and tree level. As we would expect, the frequency of correctly selected families increases with the sample size and decreases with the tree level. Overall, the correct family is selected most of the time, although there is room for improvement.

\begin{table}[h!]
	\centering
    \caption{Frequencies of the true copula family being selected (in \%). Results are split by sample size and tree level. Standard errors are given in brackets below.} 
\begin{tabular}{rrrrr}
\hline \hline
 $n$ & $T_1$ & $T_2$ & $T_3$ & $T_4$ \\ 
  \hline
500    &  76.8 &  63.0 &  53.5 & 43.0 \\ 
       & (0.9) & (1.2) & (1.6) & (2.2) \\ 
5\,000 &  97.5 &  96.7 &  87.9 & 72.2 \\ 
       & (0.3) & (0.5) & (1.0) & (2.0) \\ 
   \hline
\end{tabular}
\label{sec:simul:model_selection_1}
\end{table}

For $n=500$, \autoref{sec:simul:fams_tab1} and \autoref{sec:simul:fams_tab2} are contingency tables with the frequency for each family to be chosen correctly. In the first tree (\autoref{sec:simul:fams_tab1}), the family selection works very well for the Student $t$ copula and reasonably well for all other families. For example, when the true copula family is Gaussian, the Student $t$ copula is selected in $4.8 / 24.1 \approx 20\%$ of the cases. In the fourth tree (\autoref{sec:simul:fams_tab2}), the performance deteriorates, and the Student $t$ copula is more than 25\% of the time when the true copula is Gaussian. Similarly, when the true copula is Archimedean, elliptical copulas are selected roughly 30\% ($T_1$) and 70\% ($T_4$) of the time. Although not shown here, both effects decrease with the sample size.

\begin{table}[h!]
\centering
\caption{Contingency table with frequencies for each family to be selected for $n = 500$ and tree $T_1$ (in \%).}
\begin{tabular}{rrrrrr}
\hline \hline
\multicolumn{1}{c}{} & \multicolumn{4}{c}{true family} & \multicolumn{1}{c}{}\\
  \cline{2-5}
selected family & Gaussian & Student $t$ & Clayton & Gumbel & $\sum$ \\ 
\hline
Gaussian    & 18.2 & 0.7  & 4.3  & 5.1  & 28.4 \\ 
Student $t$ & 4.8  & 23.4 & 2.9  & 3.4  & 34.4 \\ 
Clayton     & 0.1  & 0.0  & 18.5 & 0.0  & 18.6 \\ 
Gumbel      & 1.0  & 0.9  & 0.1  & 16.7 & 18.7 \\ 
$\sum$      & 24.1 & 25.0 & 25.9 & 25.2 & 100.0 \\  
\hline
\end{tabular}
\label{sec:simul:fams_tab1}

\end{table}

\begin{table}[h!]
\centering
\caption{Contingency table with frequencies for each family to be selected for $n = 500$ and tree $T_4$ (in \%).}
\begin{tabular}{rrrrrr}
\hline \hline
\multicolumn{1}{c}{} & \multicolumn{4}{c}{true family} & \multicolumn{1}{c}{}\\
  \cline{2-5}
selected family & Gaussian & Student $t$ & Clayton & Gumbel & $\sum$ \\ 
\hline
Gaussian    & 15.0 & 5.2  & 12.0 & 12.2 & 44.4 \\ 
Student $t$ & 6.4  & 14.6 & 7.2  & 6.2  & 34.4 \\ 
Clayton     & 1.4  & 0.6  & 4.6  & 0.2  & 6.8 \\ 
Gumbel      & 2.8  & 2.2  & 0.6  & 8.8  & 14.4 \\ 
$\sum$      & 25.6 & 22.6 & 24.4 & 27.4 & 100.0 \\ 
\hline
\end{tabular}
   \label{sec:simul:fams_tab2}
\end{table}

Recall that the family is chosen based on the AIC, a trade-off between goodness of fit and model complexity. In a finite sample, the likelihood of the Student $t$ copula is necessarily larger than that of the Gaussian copula, because the former nests the latter. On the other hand, the AIC puts a larger penalty on the Student $t$ copula for the additional parameter. However, the GAM for a single pair-copula usually has more than ten parameters to account for the covariate effects. Hence, the increase in penalty resulting from selecting Student $t$ copula over the Gaussian is relatively small. This is only a minor problem, since the two families are very similar overall. As for the selection of the Student $t$ copula when the true family is Archimedean, the true model for each pair-copula is specified for the Kendall's $\tau$. Hence, while some distributional features such as Archimedean's tail asymmetries cannot be reproduced by the Student $t$ copula, the covariate effects on Kendall's $\tau$ can.

In the supplementary material we show all results when the family is selected by BIC instead of AIC. While the estimation accuracy is largely unaffected, the criterion has a notable influence on the selected families. The AIC selects the correct family more often, but the effect on individual families is rather complex and difficult to interpret.

\underline{Covariate Selection.}

\noindent Finally, we investigate the automatic selection of the covariates. \autoref{sec:simul:model_selection_2} shows the frequency that a covariate was correctly included or excluded from the model (averaged over all eleven covariates). For the choice of covariates, a correct selection means that $z_1$, $z_2$, $z_3$, $z_6$, $z_7$, and $z_8$ (linear terms), as well as $t_1$, $t_2$, and $t_3$ (smooth terms) are included, and that $z_4$, $z_5$, $z_9$, $z_{10}$, $t_4$, and $t_5$ are not. Recall from \autoref{subsec:ems_pc} that the covariates are selected out of the model based on a significance test with a $p$-value of 5\% as the threshold. We observe that the performance decreases with the tree level (e.g., from $83.5\%$ in the first tree to $54.8\%$ in the fourth for $n = 500$), but the effect is negligible for the larger sample size. Further, the performance increases with the sample size. Generally, the linear terms are correctly selected out more often, owing to the more complicated selection procedure for smooth terms.

\begin{table}[h!]
\centering
\caption{Frequencies of the correct choice of covariates (in \%). Results are split by sample size, tree level, and whether the covariate is part of the linear or smooth component. Standard errors are given in brackets below.} 
\begin{tabular}{rrrrrrrrr}
\hline \hline
  & \multicolumn{2}{c}{$T_1$} & \multicolumn{2}{c}{$T_2$} &\multicolumn{2}{c}{$T_3$} & \multicolumn{2}{c}{$T_4$} \\ 
 $n$ & linear & smooth & linear & smooth & linear & smooth & linear & smooth \\
\hline
 500   &  83.5 & 75.4 & 81.3 & 72.5 & 73.6 & 63.5 & 61.4 & 54.8 \\ 
       &  (0.3) & (0.4) & (0.3) & (0.5) & (0.4) & (0.7) & (0.7) & (1.0) \\ 
5\,000 &  94.6 & 91.5 & 94.8 & 91.4 & 94.8 & 91.2 & 94.4 & 90.2 \\ 
       &  (0.2) & (0.3) & (0.2) & (0.3) & (0.2) & (0.4) & (0.3) & (0.6) \\  
\hline
\end{tabular}
 \label{sec:simul:model_selection_2}
\end{table}

The selection algorithm also decides whether the covariates $t_j$, $j = 1, \dots, 5$ are to be included as linear or smooth components. Recall that $s_1$ is actually a linear function and, thus, should be selected as a linear term. \autoref{sec:simul:model_selection_3} shows the frequencies of this happening. The frequencies are generally larger for lower trees and larger samples. In the first tree, $t_1$ is treated as a linear term 40\% and 60\% of the time respectively, which leaves plenty of room for improvement. But note that, even when $t_1$ is not treated as a linear term, its effect can still be estimated consistently (see sections on estimation accuracy).

\begin{table}[h!]
\centering
\caption{Frequencies of the correctly selecting $t_1$ as a linear covariate (in \%). Results are split by sample size and tree level. Standard errors are less than 0.1\% in all cases.} 
\begin{tabular}{rrrrrr}
\hline \hline
 $n$ & $T_1$ & $T_2$ & $T_3$ & $T_4$ \\ 
  \hline
 500 & 42.5 & 41.9 & 31.8 & 19.2 \\ 
5000 & 58.8 & 57.6 & 58.5 & 52.2 \\ 
   \hline
\end{tabular}
 \label{sec:simul:model_selection_3}
\end{table} 
 
\underline{Computation times.}

We close the analysis by a brief discussion of computation time. \autoref{sec:simul:times} reports the time required by the two estimators to fit the full model described in \autoref{sec:simul:setup}. Recall that the oracle estimator knows the full model specification in advance; the selection estimator needs to try different specifications for the copula families, covariates, and basis size. This results in computation times that are roughly 30 times slower compared with the oracle estimator. Since we are selecting from six copula families, fitting a model for each family roughly increases the computing time by a factor of six. The remaining difference (a factor of about 5) is due to the selection of covariates and basis sizes. The magnitude of this factor is mainly driven by the number of nonparametric components. Fitting the same model with a single nonparametric term results in a factor of about two:

\begin{table}[ht]
\caption{Average computation times for estimating the GAM vine copula model (in minutes), recorded on a single thread of a 8-way Opteron (Dual-Core, 2.6 GHz) CPU with 64GB RAM. Standard deviations are shown in brackets below.}
\label{sec:simul:times}
\centering
\begin{tabular}{rrr}
  \hline \hline
 n & oracle & selection \\ 
  \hline
500    &  0.4  &  12.8 \\ 
       & (0.1) & (1.2) \\
5\,000 &  2.0  &  68.8 \\
       & (0.5) & (6.4) \\
   \hline
\end{tabular}

\end{table}

The computing time for the selection estimator may seem large. But recall that we are estimating a model with 370 parameters and need to select the model structure and copula family for all 10 pair-copula families. Furthermore, there are ways to drastically reduce the computing time without giving up much in performance. 
\begin{enumerate}
\item The estimation and model selection for individual pair-copulas can be parallelized within each tree-level. Recall that there are $d - k$ pair-copulas in tree $T_k$. As a rule of thumb, we can expect the computing time to be reduced by a factor of roughly $(d-1)/2$ (assuming a sufficient number of cores).
\item For each pair-copula, we can select the covariates and basis size only for one family and use the same model structure for all other families. Using a total of six families, this reduces the computing time by a factor of four. As shown in the supplementary material, this only has a very small effect on overall performance.
\end{enumerate}

\section{Application}
\label{sec:appli}
We use the methodology developed in this paper to study the cross-sectional dynamics of intraday asset returns. They offer a magnifying glass to study financial markets while posing unprecedented econometric challenges. More specifically, we focus on the foreign exchange (FX) market, which determines the relative value of currencies. The two main characteristics of this decentralized market are that it operates both around the clock (from Sunday 10pm to Friday 10pm UTC) and around the globe (i.e., it is geographically dispersed). Recently, \cite{vatterchavez2015} observed that the intraday dependence structure pattern, due to the cyclical nature of market activity, is shaped similarly to that of the univariate conditional second moments. 

In what follows, we extend their bivariate model to encompass three or more exchange rates using vines with exogenous covariates. We use data graciously provided by Dukascopy Bank SA (\url{www.dukascopy.com}), an electronic broker holding a Securities Dealer License issued by the FINMA. It contains 15-minute spaced returns (i.e., 96 observations each day) for the EURUSD, GBPUSD, USDCHF, and USDJPY, from March 10, 2013 to November 1, 2013. Hence, in a total of 34 trading weeks, there are 16320 observations (170 days) excluding weekends. In the following, we model exchange rates from a purely time-series perspective without economic covariates, namely using only past-values and time as covariates.

The R code for this analysis is available as supplementary material.

\subsection{Modeling the Marginal Distributions}
Because intraday returns are heteroskedastic, we need to pre-filter the individual series before applying the methodology of this paper. In this context, the high-frequency econometrics literature usually decomposes the volatility in two multiplicative components: a seasonal, but often assumed deterministic, and a stochastic part (see e.g., \citealt{andersen_bollerslev:1997,andersen_bollerslev:1998,engle_sokalska:2012}). It is straightforward to achieve this within the GARCH-family, where $r_t = \sigma_t \, y_t$ for $\sigma_t$ a function of $\left\{r_{t-1}, \sigma_{t-1}, r_{t-2}, \sigma_{t-2}, \cdots \right\}$ and $y_t$ a white noise, which we do by writing

\begin{align*}
\log \sigma_t^2 = &\left[ \omega + \sum^K_{k=1} \left\{a_k\, \cos\left(2\pi k  t/T \right)+b_{k}\, \sin\left(2\pi  k t/T \right)\right\} \right] \\ 
&+ \alpha \, \epsilon_{t-1} + \gamma \left(\mid \epsilon_{t-1} \mid - \mathbb{E} \mid \epsilon_{t-1} \mid\right) + \beta \log \sigma^2_{t-1},
\end{align*}
where $T=96$. This model is the EGARCH(1,1) from \cite{nelson1991}, augmented with external regressors to take the seasonality into account.  The sum of cosines and sines with integer frequencies, designed to capture daily oscillations around the base level, is similar to the Fourier Flexible Form (FFF, see \citealt{gallant:1981}), introduced in this context by \cite{andersen_bollerslev:1997,andersen_bollerslev:1998}.
Denoting by $\widehat{\sigma}_t$ the fitted volatility using the maximum log-likelihood estimator with $K=5$, we call $\widehat{y}_t = r_t\, /\, \widehat{\sigma}_t$ the residuals.

\begin{figure}
    \vspace{-0.5cm}
    \centering
    \subfloat[EURUSD]{
	\includegraphics[width=0.81\textwidth]{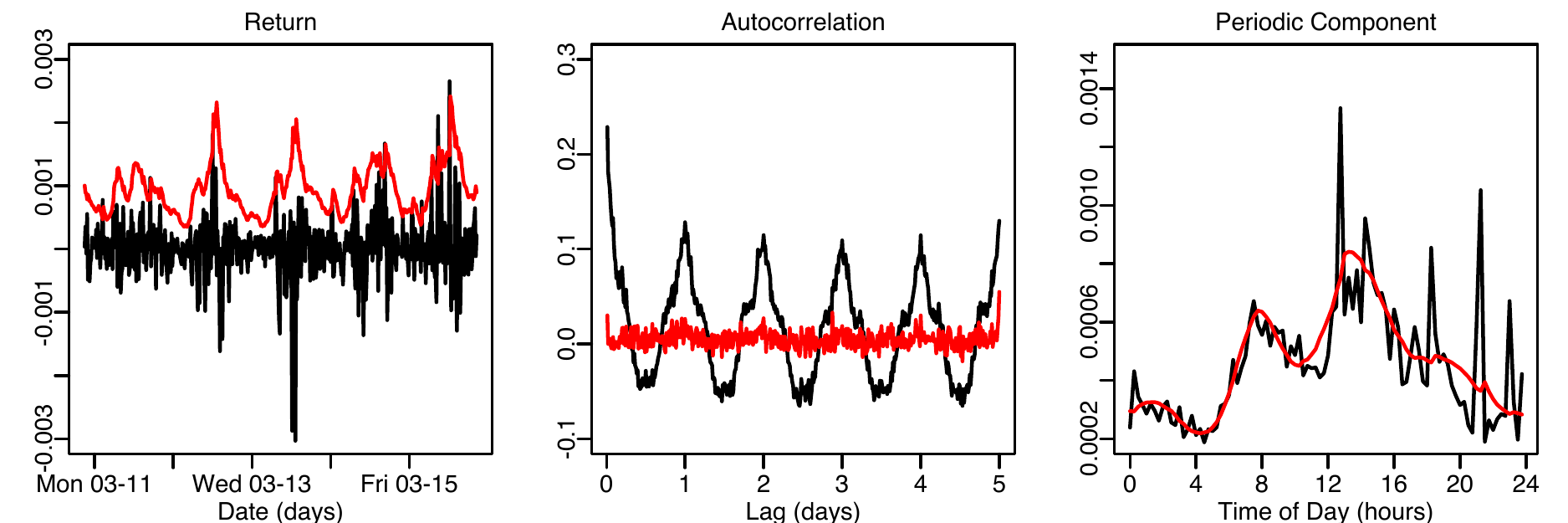}
    } \\
    \subfloat[GBPUSD]{
        \includegraphics[width=0.81\textwidth]{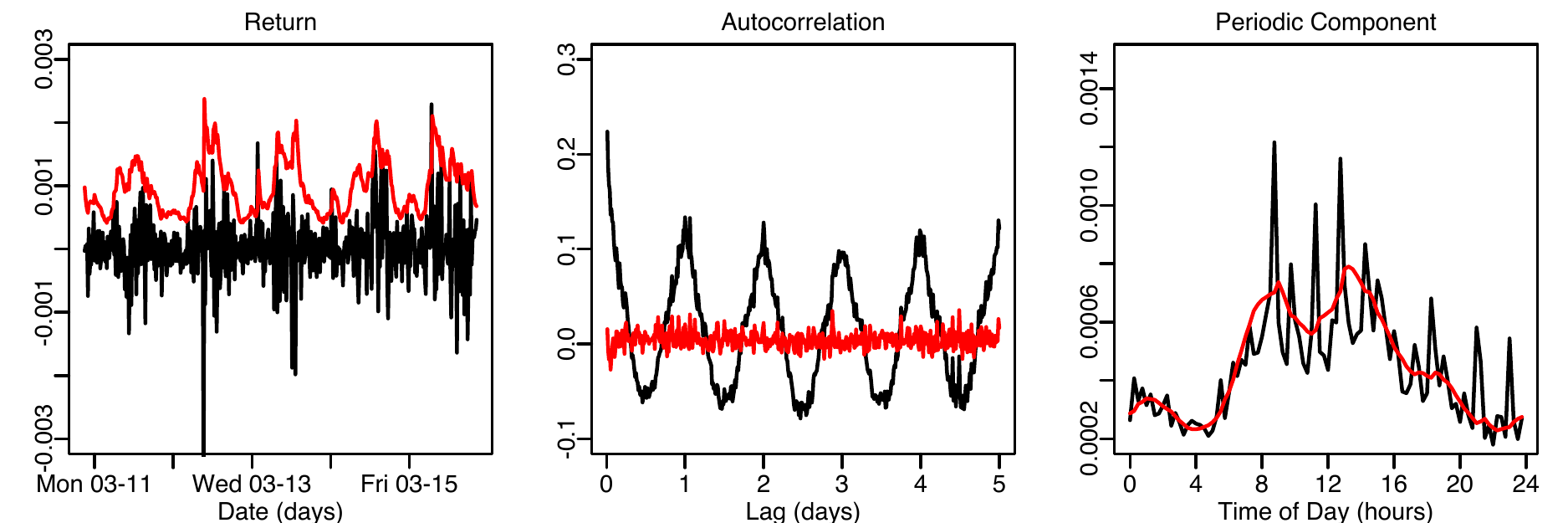}
    } \\
    \subfloat[USDCHF]{
    \includegraphics[width=0.81\textwidth]{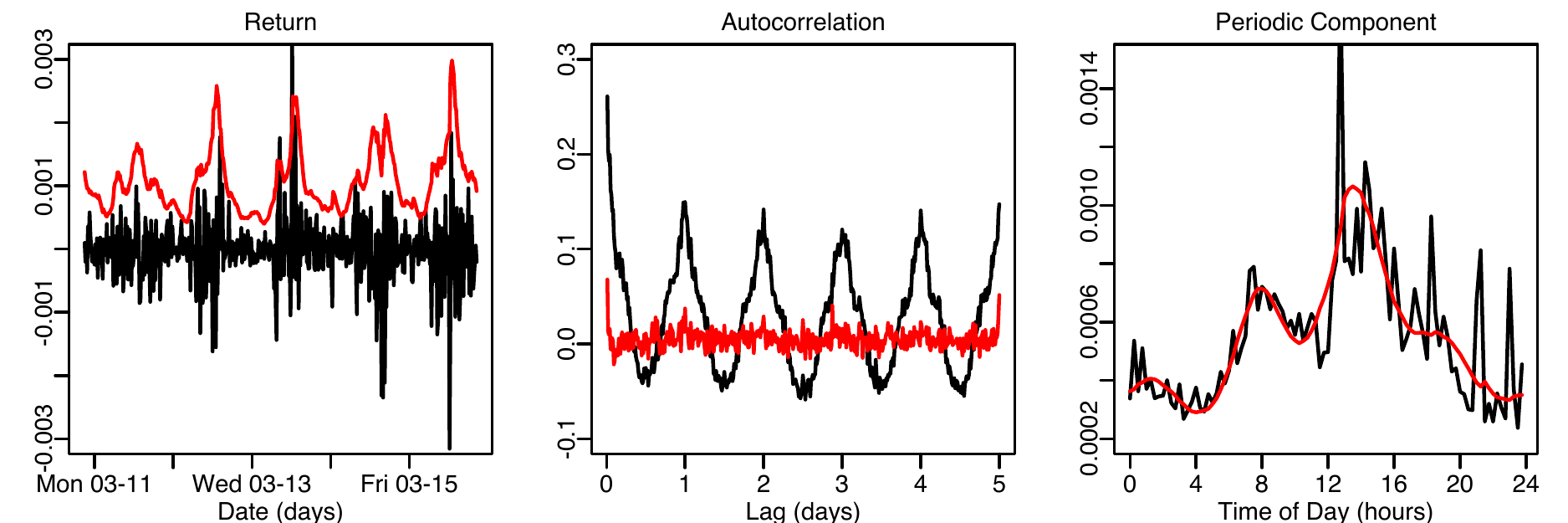}
    } \\
    \subfloat[USDJPY]{
       \includegraphics[width=0.81\textwidth]{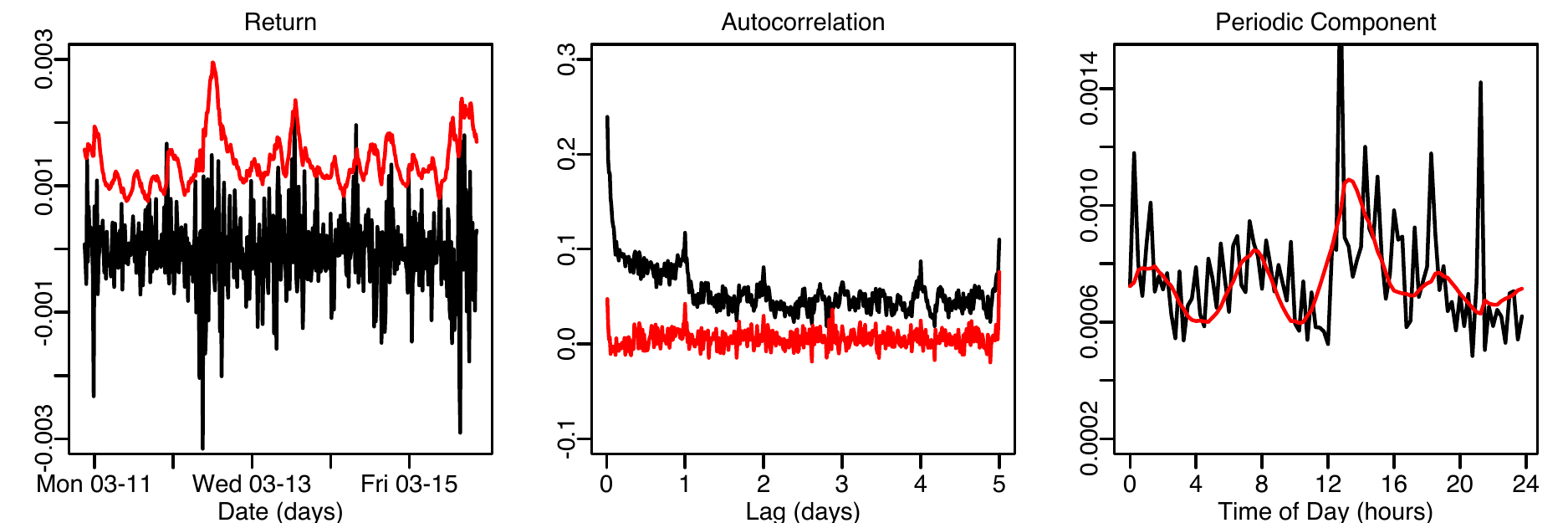}
    }
	\caption{Marginal modeling of the Four FX rates. In the left panels, the return, $r_t$, (black line) and two conditional standard deviations, $2 \times \widehat{\sigma}_t$, (red line) are shown for the first week of the sample. In the middle panels, the autocorrelation of the absolute value of the return/deseasonalized residual are the black/red lines. In the right panels, the black (respectively red) curve represents the empirical (respectively fitted)  volatility per 15-minute bin.}
	\label{sec:appli:highfreq_eurusd}
\end{figure}

In the left panels of \autoref{sec:appli:highfreq_eurusd}, we show the returns, $r_t$, along with two fitted conditional standard deviations, $2 \times \widehat{\sigma}_t$, for each exchange rate for the first week. In the middle panels of \autoref{sec:appli:highfreq_eurusd}, the black (respectively red) curve represents the autocorrelation of the absolute value of the returns (respectively residuals), where we observe that our univariate models appropriately capture the heteroskedasticity. In the right panels of  \autoref{sec:appli:highfreq_eurusd}, the black (respectively red) curve represents the empirical (respectively fitted)  volatility per 15-minute bin, where we recognize the usual modes at the opening time of the Tokyo, London and New-York markets.

\begin{figure}
	\centering{\centering
\subfloat[Tree $T_1$]{
\begin{minipage}[t]{.32\linewidth}
     \centering
\begin{tikzpicture}[node distance=2cm,>=latex']
\node [nn1] (U1) {$1$};
\node [nn1, below=1.8cm of U1] (U3) {$3$};
\draw [ee1] (U1) --  node {$1,3$} (U3) ;
\node [nn1, right = of U1] (U2) {$2$};
\draw [ee1] (U1) --  node {$1,2$} (U2) ;
\node [nn1, right  = of U3] (U4) {$4$};
\draw [ee1] (U3) --  node {$3,4$} (U4) ;
\end{tikzpicture}
\end{minipage}
\label{fig:rvineappli1}
}
\subfloat[Tree $T_2$]{
\begin{minipage}[t]{.32\linewidth}
     \centering
\begin{tikzpicture}[node distance=2cm,>=latex']
\node [nn1] (U3) {$1,3$};
\node[above= 2cm of U1](U5){};
\node [nn1, right = 1cm of U5] (U1) {$1,2$};
\draw [ee1] (U1) --  node {$2,3;1$} (U3) ;
\node [nn1, left = 1cm  of U5] (U4) {$3,4$};
\draw [ee1] (U3) --  node {$1,4;3$} (U4) ;
\end{tikzpicture}
\end{minipage} 
\label{fig:rvineappli2}
}
\subfloat[Tree $T_3$]{
\begin{minipage}[t]{.2\linewidth}
     \centering
\begin{tikzpicture}[node distance=2cm,>=latex']
\node [nn1] (U1) {$2,3;1$};
\node [nn1, below = 1.8cm of U1] (U3) {$1,4;3$};
\draw [ee1] (U1) --  node {$2,4;1,3$} (U3) ;
\end{tikzpicture}
\end{minipage}
\label{fig:rvineappli3}
} }
	\caption{The Regular Vine Tree Sequence for the Four FX rates. The numbers represent the FX rates (EURUSD: 1; GBPUSD: 2; USDCHF: 3; USDJPY: 4), $x,y$ denote the bivariate distribution of $x$ and $y$, and $x,y;z$ denote the bivariate distribution of $x$ and $y$ conditional on $z$. Each edge corresponds to a bivariate pair-copula in the PCC.}
	\label{sec:appli:RVine_fig}
\end{figure}
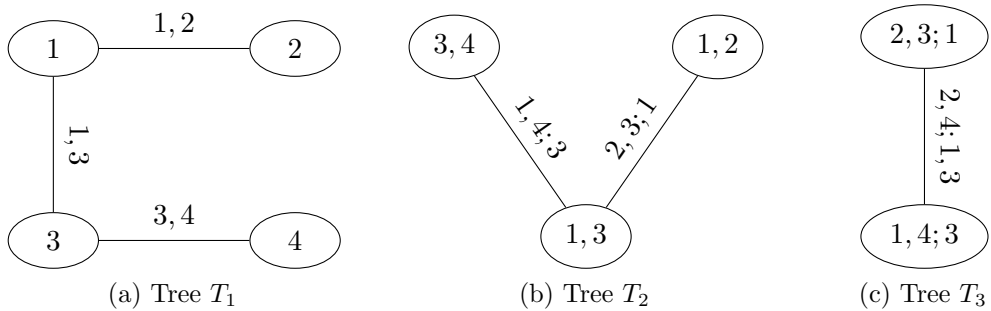

\subsection{Modeling the Dependence Structure}
From the residuals, we compute observations on the copula scale by using the empirical cumulative distribution for each individual time series. We use the same FFF regressors to model the periodic component of the dependence structure. We also add a smooth function of time $t$, to model the evolution of the dependence over the sample period. We then run the procedure described in \autoref{subsec:ems_pcc}. The resulting fitted model for the corresponding conditional pair-copulas can be written as

\begin{align*}
\tau_{1,2}(\bm w) &=  g_{1,2}\left\{ \bx(t)^{\top} \vbeta_{1,2}+s_{_{1,2}}(21.13,t)\right\}, \\
\tau_{1,3}(\bm w) &=  g_{1,3}\left\{ \bx(t)^{\top} \vbeta_{1,3}+s_{_{1,3}}(63.12,t)\right\}, \\
\tau_{3,4}(\bm w) &=  g_{3,4}\left\{ \bx(t)^{\top} \vbeta_{3,4}+s_{_{3,4}}(58.25,t)\right\}, \\
\tau_{2,3;1}(\bm w) &=  g_{2,3;1}\left\{ \bx(t)^{\top} \vbeta_{2,3;1}+s_{_{2,3;1}}(9.53,t)\right\},\\
\tau_{1,4;3}(\bm w) &=  g_{1,4;3}\left\{ \bx(t)^{\top} \vbeta_{1,4;3}+s_{_{1,4;3}}(27.96,t)\right\},\\
\tau_{2,4;1,3}(\bm w) &=  g_{2,4;1,3}\left\{ \bx(t)^{\top} \vbeta_{2,4;1,3}+s_{_{2,4;1,3}}(21.77,t)\right\},
\end{align*}
where $\bx(t) = (1, \cos\left(2\pi t/T \right), \ldots, \cos\left(2\pi 5 t/T \right),\sin\left(2\pi t/T \right), \ldots, \sin\left(2\pi 5 t/T \right))^{\top}$, $1= \mbox{EURUSD}$, $2=\mbox{GBPUSD}$, and $3=\mbox{USDCHF}$, and $4=\mbox{USDJPY}$, and the first number in each smooth function corresponds to the estimated EDF.  As is often the case with financial data, the Student $t$ copula is selected by both the AIC and BIC for all conditional pair-copulas over the Gaussian or common Archimedean copulas.

\begin{figure}
	\vspace{-0.5cm}
    \centering
    \subfloat[Smooth and Periodic Component for the GBPUSD-EURUSD copula.]{
	\includegraphics[width=0.65\textwidth]{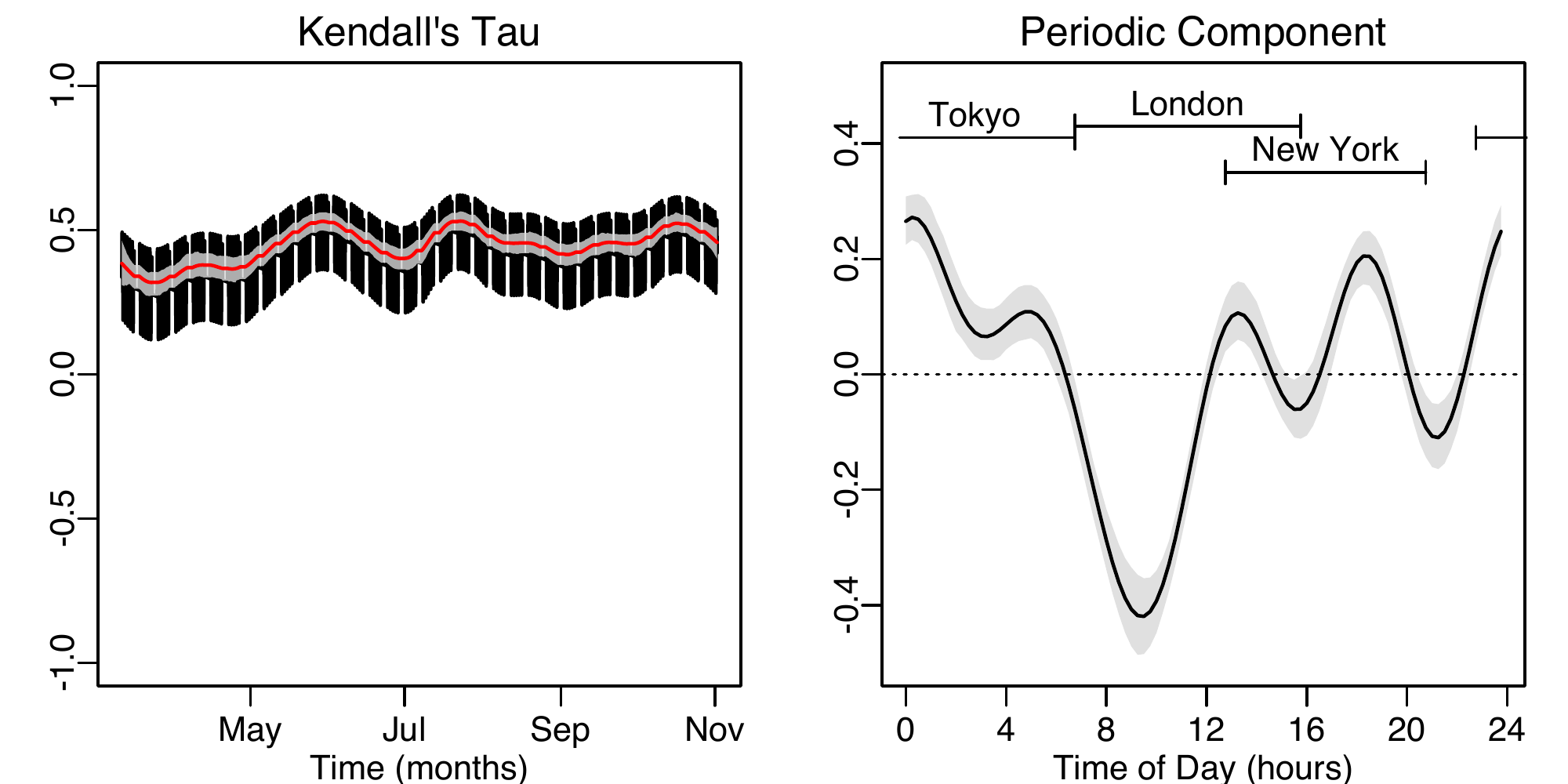}
    \label{sec:simul:highfreq_copula1}
    } \\
    \subfloat[Smooth and Periodic Component for the USDJPY-USDCHF copula.]{
	\includegraphics[width=0.65\textwidth]{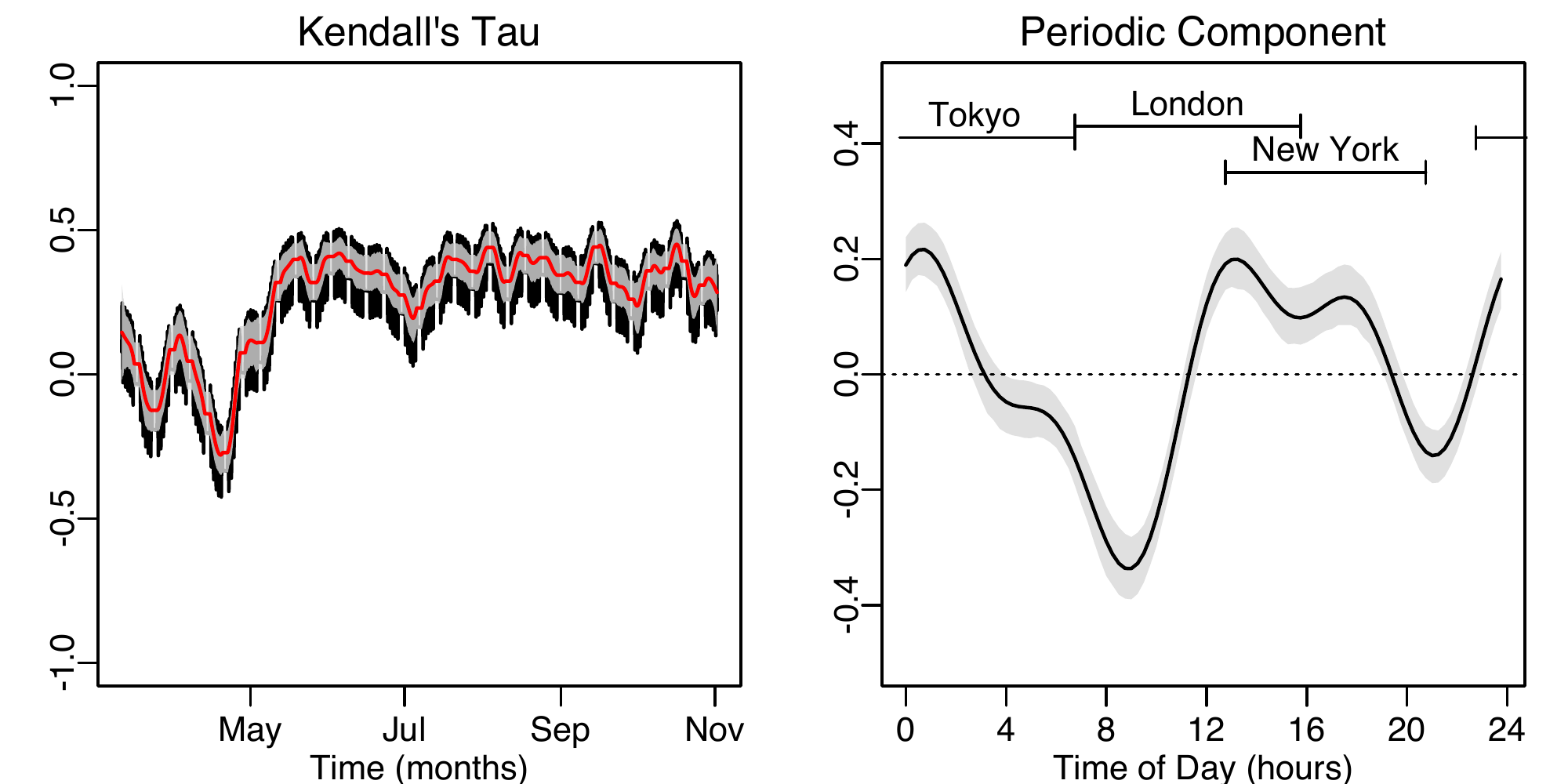}
    \label{sec:simul:highfreq_copula3}
    } \\
    \subfloat[Smooth and Periodic Component for the EURUSD-USDCHF copula.]{
	\includegraphics[width=0.65\textwidth]{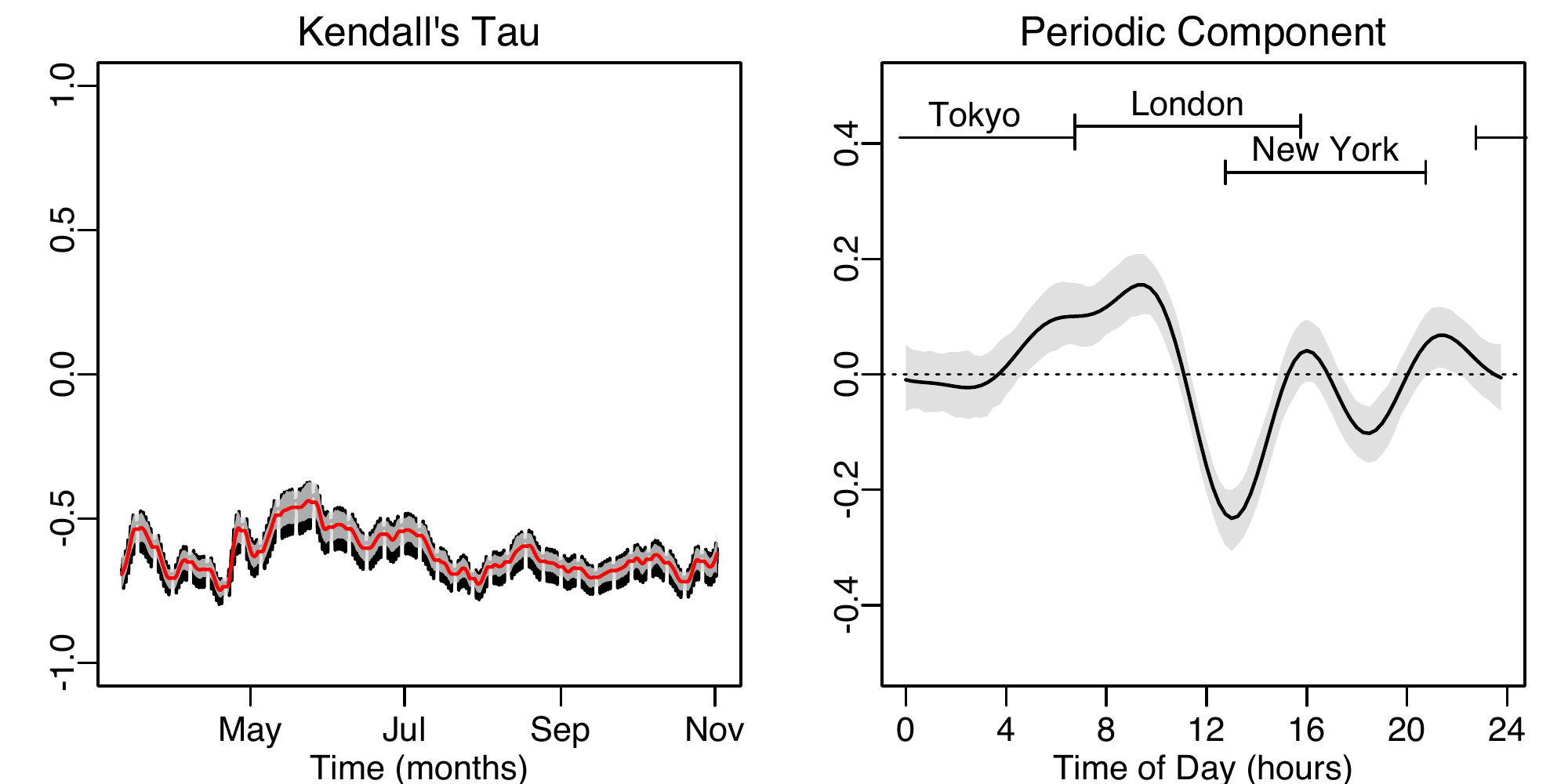}
	\label{sec:simul:highfreq_copula2}
    }

\caption{Results for the First Tree of the PCC for the Four FX rates. In the left panels, the sum of the smooth and the periodic components is the black line, and the smooth component with its 95\% confidence band is the red line with shaded grey area. All quantities are shown on the Kendall's $\tau$ scale. In the right panels, the periodic component with its 95\%  confidence band is the black line with shaded grey area. All quantities are shown on the linear predictor scale.}
	\label{sec:appli:highfreq_results}
	\vspace{-0.5cm}
\end{figure}

\begin{figure}
	\vspace{-0.5cm}
    \centering
    \subfloat[Smooth and Periodic Component for the GBPUSD-USDCHF;EURUSD copula.]{
    \includegraphics[width=0.65\textwidth]{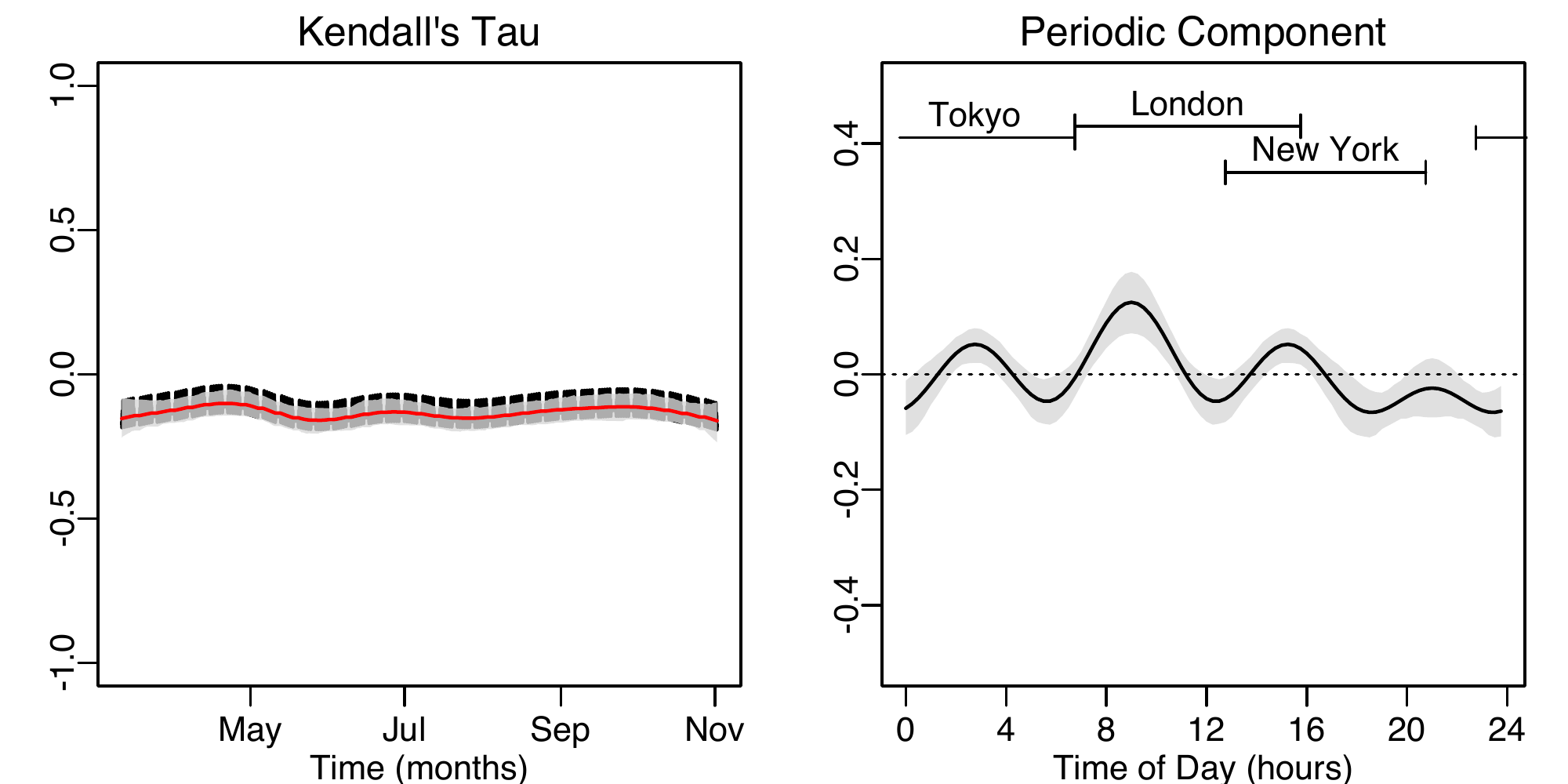}
    \label{sec:simul:highfreq_copula4}
    } \\
    \subfloat[Smooth and Periodic Component for the EURUSD-USDJPY;USDCHF copula.]{
    \includegraphics[width=0.65\textwidth]{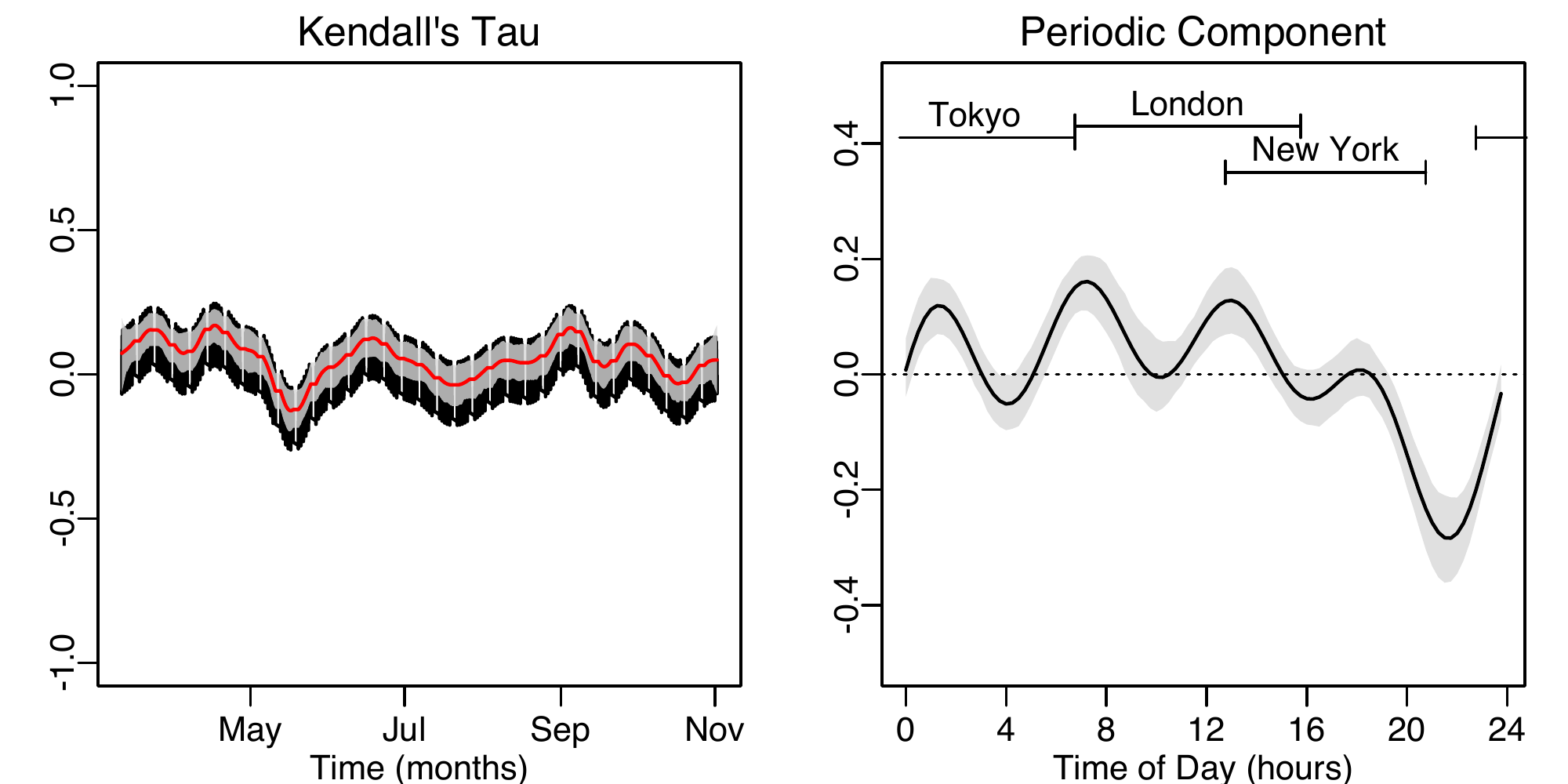}
    \label{sec:simul:highfreq_copula5}
    } \\
    \subfloat[Smooth and Periodic Component for the GBPUSD-USDJPY;USDCHF,EURUSD copulas.]{
	\includegraphics[width=0.65\textwidth]{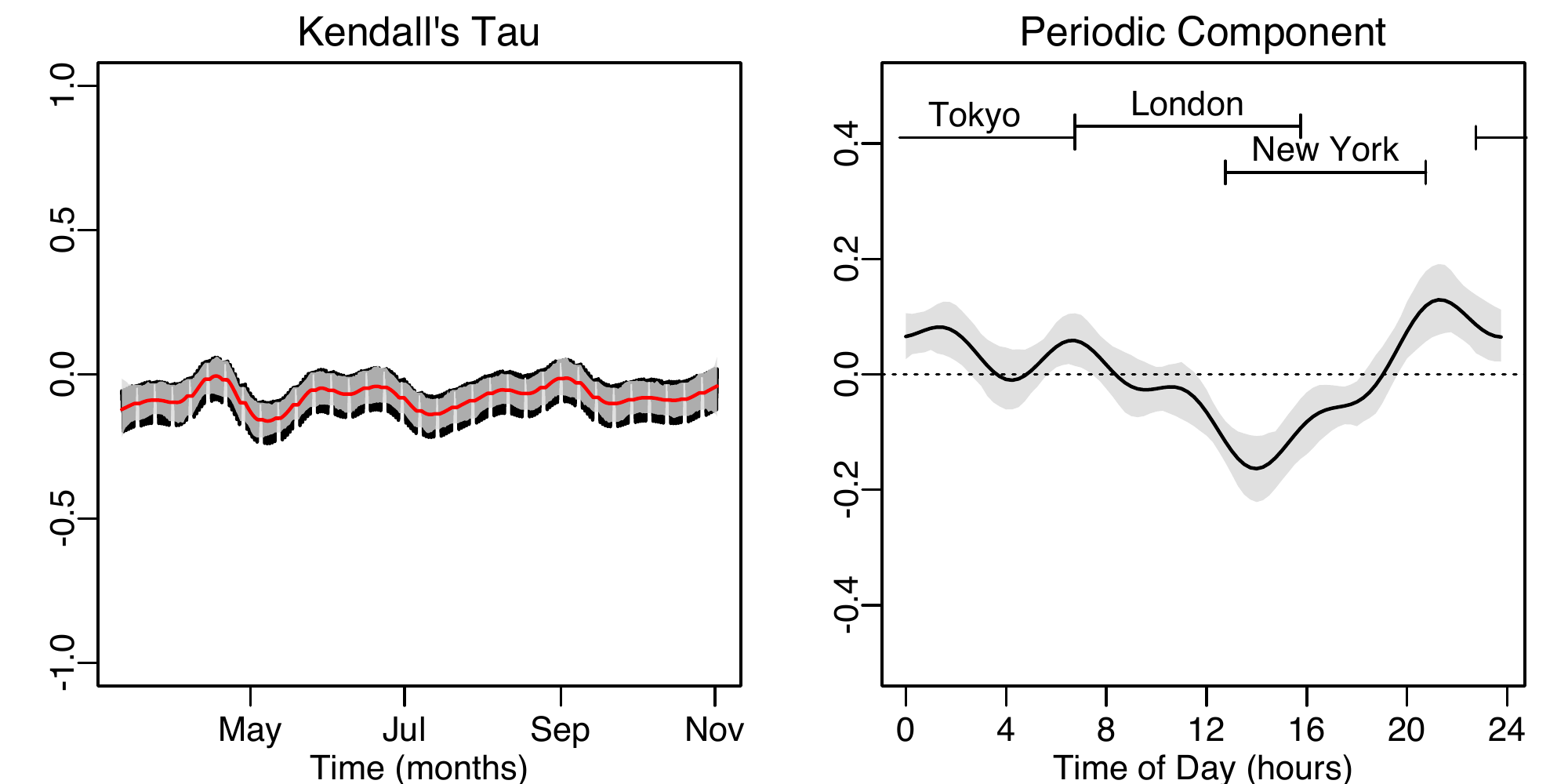}
    \label{sec:simul:highfreq_copula6}
    }
\caption{Results for the Second and Third Trees of the PCC for the Four FX rates. In the left panels, the sum of the smooth and the periodic components is the black line, and the smooth component with its 95\% confidence band is the red line with shaded grey area. All quantities are shown on the Kendall's $\tau$ scale. In the right panels, the periodic component with its 95\%  confidence band is the black line with shaded grey area. All quantities are shown on the linear predictor scale.}
	\label{sec:appli:highfreq_results2}
	\vspace{-0.5cm}
\end{figure}

In \Autoref{sec:appli:highfreq_results,sec:appli:highfreq_results2}, we show all fitted smooth and periodic components (without the intercept). In the left panels of both figures, the sum of the smooth and the periodic components, that is $g\left\{\bx(t)^{\top} \widehat{\vbeta}+\widehat{s}(t)\right\}$ is the black line, oscillating wildly because of the daily periodicity. For the sake of clarity, we also show $g\{\widehat{s}(t)\}$ with bootstrapped 95\% confidence bands as the red line with shaded grey area. In this case, we assume that $\bx(t)^{\top} \widehat{\vbeta} = 0$, which is sensible since the periodic component averages zero over one day. Finally, in the right panels, we show the periodic component only (black line), on the scale of the linear predictor, that is $\bx(t)^{\top} \widehat{\vbeta}$, with bootstrapped 95\% confidence bands (shaded grey area). 

\underline{Dependence Analysis for the First Tree.}

\noindent In the first tree (i.e., the first three pair-copulas), a pair-copula describes directly the time-varying dependence between two currency pairs. For instance, the red curve in each of the left panels of \Autoref{sec:simul:highfreq_copula1,sec:simul:highfreq_copula3,sec:simul:highfreq_copula2}  illustrates the long-term evolution of Kendall's $\tau$. We observe that the dependence is positive for GBPUSD-EURUSD and USDJPY-USDCHF, but negative for EURUSD-USDCHF. Similarly, the patterns in the right panels of \autoref{sec:simul:highfreq_copula2} essentially mirror the one of \Autoref{sec:simul:highfreq_copula1,sec:simul:highfreq_copula3}. The explanation lies in the position of the USD in the currency pairs. In the GBPUSD-EURUSD, the USD is the second leg of both rates, while in the USDJPY-USDCHF, the USD is the first leg of both rates. Conversely, for the EURUSD-USDCHF, the USD appears as the first and second leg of each exchange rate. The absolute valuation of the USD is mainly related to the health of the US economy. Hence, when a piece of information which the market interprets as positive for the US economy is released (e.g., a diminishing unemployment rate), the USD becomes more valuable. When this is the case, the GBPUSD and EURUSD both decrease, and the USDJPY and USDCHF both increase. Since, in the GBPUSD-EURUSD and USDJPY-USDCHF, the USD is priced by the two legs in the same direction, the dependence is positive. Conversely, in the EURUSD-USDCHF, the USD is priced by the two legs in opposite directions, and the dependence is negative.

For the right panels of \Autoref{sec:simul:highfreq_copula1,sec:simul:highfreq_copula3,sec:simul:highfreq_copula2}, we see two local peaks at the opening and closing of the US market. For the GBPUSD-EURUSD and USDJPY-USDCHF, respectively the EURUSD-USDCHF, the peaks are positive, respectively negative, that is of the same sign as the overall dependence. In other word, the absolute dependence is increased at the opening and closing of the US market, which represents evidence that the USD's valuation is a driver of the dependence between exchange rates. Comparatively, the absolute dependence becomes smaller when the London market opens or closes. This is interesting, because it contradicts the pattern observed in the volatility, where peaks were observed at the openings of the Asian, European, and US markets.


\underline{Dependence Analysis for the Subsequent Trees.}

\noindent In the second and third trees, a pair-copula describes the dependence between two currency pairs after the effect of other currency pairs has been removed. For example, the pair copula for EURUSD-USDJPY;USDCHF is the residual dependence between EURUSD and USDJPY, once the dependence induced by the USDCHF has been removed. The USD always appears in one leg of the currency pairs whose influence is removed. Hence, we expect the USD to be less influential in the second and third trees.

Firstly, we observe that the long-term evolution become less relevant. Not only the red curves in the left panels of \Autoref{sec:simul:highfreq_copula4,sec:simul:highfreq_copula5,sec:simul:highfreq_copula6} are much smoother than in those of \Autoref{sec:simul:highfreq_copula1,sec:simul:highfreq_copula3,sec:simul:highfreq_copula2}, they are also much closer to zero. 
Comparing \Autoref{sec:simul:highfreq_copula4,sec:simul:highfreq_copula5}, the conditional pair involving the GBPUSD is less wiggly than the one involving the USDJPY. The explanation comes from \Autoref{sec:simul:highfreq_copula1,sec:simul:highfreq_copula3}, where the same effect is observed: the pair involving the USDJPY (i.e., the USDJPY-USDCHF) is more wiggly than the one involving the GBPUSD (i.e., the GBPUSD-EURUSD).

Secondly, we still see significant periodic patterns in all pairs. For instance, in the right panel of \autoref{sec:simul:highfreq_copula4} (i.e., the EURUSD-USDJPY;USDCHF pair-copula), we observe positive dependence peaks at opening and closing times of the Tokyo and London markets. So after the influence of the USDCHF pair has been removed, the EURUSD and USDJPY appear to be positively dependent when the European and Japanese markets dominate the trading. Conversely, when both the London and Tokyo markets are closed, the dependence becomes more negative, which is likely caused by the opposite positioning of the USD in each pair. In \autoref{sec:simul:highfreq_copula5} (i.e., the GBPUSD-USDCHF;EURUSD pair-copula), we see two small peaks at the opening and closing of the London market, suggesting that information about the European economies induces positive dependence between GBPUSD and USDCHF. Lastly, the right panel of \autoref{sec:simul:highfreq_copula6} (i.e., the GBPUSD-USDJPY;USDCHF,EURUSD pair-copula) exhibits positive dependence whenever neither London nor New York is trading, and negative dependence during New York's opening hours. However, such a complex (second-order and non-linear) dependence relationship is rather difficult to interpret.

\underline{Summary.}

\noindent In summary, we found strong evidence for a dynamic dependence structure in intraday foreign exchange rates. Our analysis suggests that it is appropriate to decompose the time-varying dependence into two components. The first captures the long-term evolution of the dependence. The second captures the daily patterns stemming from the periodic nature of the market, which is related to the opening and closing times of various exchanges around the world.  This analysis is similar to the well-known patterns in the intraday volatility observed in the individual returns on intraday foreign exchange rates. 

\section{Discussion}
\label{sec:discussion}
In this paper, we extend pair-copula constructions (PCCs) by letting the Kendall's tau of each pair-copula be a function of covariates. We utilize the flexibility of generalized additive models (GAMs), which allow for parametric, semi-parametric or non-parametric specifications of the relationship between strength of dependence and the covariates. Building on the maximum penalized log-likelihood estimator for conditional copulas of \cite{vatterchavez2015}, we propose a sequential estimation algorithm, as well as a heuristic method for a fully automatic model selection. We evaluate both in a simulation study, and we find that 
\begin{itemize}
\item the estimates are unbiased in the first tree, but there is a shrinkage towards zero of increasing size in subsequent trees, 
\item the performance of the selection and estimation are comparable to that of an oracle estimator (where the structure and copula families are known),
\item the model selection heuristic selects the true copula family and covariates most of the time, but the frequency gets smaller with increasing tree level.
\end{itemize}

We used this methodology to model the dependence between intraday returns of four exchange rates. We observed that the bivariate results of \cite{vatterchavez2015} extend directly in this higher-dimensional example. In other words, the data suggest that the dependence can be decomposed into a smooth and a periodic component. Furthermore, the periodic component for each conditional pair-copula has peaks that correspond to openings and/or closings of markets around the world. While most of the time-varying features are captured in the first tree, there is still a significant amount of periodicity left in the second and third trees of the PCC model.

The simulations and application presented in this paper feature only a medium number of covariates. When the number of covariates grows, the algorithms that we use may be slow and run into numerical difficulties. As such, an inherent limitation of our method is its inability to handle high dimensional covariates. To overcome this difficulty, there are (at least) three potential directions, which, although promising, are out of the scope of this paper. First, we could use alternative sparsity-enforcing penalties as in \citet{Chouldechova2015,Lou2016,Petersen2016}. Second, we could explore Boosting-related ideas as in \citet{Buhlmann2003,Buhlmann2007,Tutz2007,Schmid2008}. Third, we could apply a dimensionality reduction technique to the covariates. 

This paper represents, to the best of our knowledge, the first attempt at modeling copulas in more than three dimensions conditionally on more than one covariate. While there exist numerous copula families in the bivariate case, the options in higher-dimensions are rather limited. This has inspired the development of hierarchical models, constructed from cascades of simpler building blocks. While PCCs are a class of such models, factor copulas as in \cite{krupskiijoe2013,krupskiijoe2015} define another. Thanks to their appealing computational properties, the later represent a promising alternative to let copulas be functions of covariates. Although out of the scope of this paper, this approach is currently under investigation.

\section*{Acknowledgements}
The authors would like to thank for useful comments and discussions Val\'erie Chavez-Demoulin, Claudia Czado, and Anthony Davison, as well as participants from the 2016 conference on Dependence Modeling in Finance, Insurance and Environmental Science in Munich.
\pagebreak
\section*{Appendix: Model Selection}
\label{sec:appendix}
In \autoref{algo1}, $PMLE$ is a function that computes $\pmle$ and selects $\vgamma$ by GVC minimization as in \citet{vatterchavez2015}. Its inputs are two response vectors, two matrices of parametric and nonparametric covariates and a vector of basis sizes. Its outputs are the fitted model, along with the $p$-values for each covariate and the EDFs for the smooth components. 

\begin{algorithm}
\caption{Model selection for a bivariate conditional copula with known family}
  \label{algo1}
  \begin{scriptsize}
  \begin{algorithmic}[1]
  \Inputs{$u_1,u_2,x_1,\ldots,x_k,\alpha$}
    \Initialize{\strut $b \gets k$ \\
     $basis_j \gets 10$, $j=1, \ldots ,b$ \\
    $cov \gets$ a $n \times b$ matrix with columns $x_1,\ldots,x_b$\\
    $lincov \gets$ an empty matrix \\
    $sel \gets false$}
    \While{ANY$(sel\not=true)$ AND $b >0$} 
    \Comment Remove ``insignificant'' covariates and determine linear covariates
    \State $fitted \gets$ PMLE$(u_1,u_2,lincov,cov,basis)$
    \State Clear $basis$, $PV$, and $sel$
    \State $PV_j \gets$  $p$-value of $fitted$ corresponding to column $j$ of $cov$, $j=1,\ldots,b$
    \State $sel_j \gets true$, $j=1,\ldots,b$
    \For{$j = 1$ to $b$}
    \If{$PV_j \geq \alpha$} 
    \State $sel_j \gets false$ 
    \State remove column $j$ from $cov$
    \EndIf
    \If{$EDF_j \leq 1.5$} 
    \State $sel_j \gets false$ 
    \State add column $j$ from $cov$ to $lincov$
    \State remove column $j$ from $cov$
    \EndIf
    \EndFor 
	\State $b \gets \sum^b_{j = 1} \ind_{PV_j < \alpha \mbox{AND} EDF_j > 1.5}$
    \State $basis_j \gets  10$, $j=1, \ldots ,b$
    \EndWhile
    \State $fitted \gets$ PMLE$(u_1,u_2,lincov,cov,basis)$
    \If{$b \not= 0$}
    \Comment Select the basis size
    \State $EDF_j \gets$ EDF of $fitted$ corresponding to column $j$ of $cov$, $j=1,\ldots,b$
      \While{ANY$(sel==true)$ AND $ALL(basis < n/30)$}
      \State Clear $sel$
    \State $sel_j \gets false$, $j=1,\ldots,b$
     \For{$j = 1$ to $b$}
    \If{$EDF_j > basis \times 0.8$ AND $2 \cdot basis_j >$  number of unique values in column $j$ of $cov$} 
    \State $sel_j \gets true$
    \State $basis_j \gets 2 \cdot basis_j $ 
    \EndIf
    \EndFor
    \State $fitted \gets$ PMLE$(u_1,u_2,lincov,cov,basis)$
    \State $EDF_j \gets$ EDF of $fitted$ corresponding to column $j$ of $cov$, $j=1,\ldots,b$
    \EndWhile
	\EndIf
     \State \Return $fitted$
  \end{algorithmic}
  \end{scriptsize}
\end{algorithm}

\newpage

\section*{Supplementary Material}
\begin{description}
\item[R-package \texttt{gamCopula}:]  \quad \\[-18pt]
\begin{center}
\url{https://cran.r-project.org/web/packages/gamCopula/}
\end{center}
The package contains various tools to apply generalized additive models to bivariate copulas and PCCs, including functions for parameter estimation, model selection, simulation, and visualization.
\item[R scripts (with results):]\quad \\[-18pt]
\begin{itemize}
\item to reproduce the results in this article:
\begin{center}
\url{https://gist.github.com/tnagler/ea8190914c44f9ac184ddd962cad75f7}
\end{center}
\item  additional simulations using the BIC as criterion for selection the copula famil
\begin{center}
\url{https://gist.github.com/tnagler/457ce0bf10f9312dfc0545f1854c7afa}
\end{center}
\item additional simulations selecting the GAM structure only for one family:
\begin{center}
\url{https://gist.github.com/tnagler/48ffd0274022601daf99b64f55f8bc94}
\end{center}
\item code for the intraday FX application (without data):
\begin{center}
\url{https://gist.github.com/tnagler/73044bb254ff51f388fb67bbe5b8f28c}
\end{center}
\end{itemize}
\end{description}

\bibliographystyle{agsm}
\bibliography{tex/mybib}

\end{document}